# Pluto's Haze as a Surface Material

W.M. Grundy[1], T. Bertrand[2], R.P. Binzel[3], M.W. Buie[4], B.J. Buratti[5], A.F. Cheng[6], J.C. Cook[7], D.P. Cruikshank[2], S.L. Devins[5], C.M. Dalle Ore[2,8], A.M. Earle[3], K. Ennico[2], F. Forget[9], P. Gao[10], G.R. Gladstone[11], C.J.A. Howett[4], D.E. Jennings[12], J.A. Kammer[11], T.R. Lauer[13], I.R. Linscott[14], C.M. Lisse[6], A.W. Lunsford[12], W.B. McKinnon[15], C.B. Olkin[4], A.H. Parker[4], S. Protopapa[4], E. Quirico[17], D.C. Reuter[12], B. Schmitt[16], K.N. Singer[4], J.A. Spencer[4], S.A. Stern[4], D.F. Strobel[17], M.E. Summers[18], H.A. Weaver[6], G.E. Weigle II[11], M.L. Wong[10], E.F. Young[4], L.A. Young[4], and X. Zhang[19]

1. Lowell Observatory, Flagstaff AZ
2. NASA Ames Research Center, Moffett Field CA
3. Massachusetts Institute of Technology, Cambridge MA
4. Southwest Research Institute, Boulder CO
5. NASA Jet Propulsion Laboratory, La Cañada Flintridge CA
6. Johns Hopkins University Applied Physics Laboratory, Laurel MD
7. Pinhead Institute, Telluride CO
8. SETI Institute, Mountain View CA
9. Laboratoire de Météorologie Dynamique (CNRS/UPMC), Paris France
10. California Institute of Technology, Pasadena CA
11. Southwest Research Institute, San Antonio TX
12. NASA Goddard Space Flight Center, Greenbelt MD
13. National Optical Astronomy Observatory, Tucson AZ
14. Stanford University, Stanford CA
15. Washington University of St. Louis, St. Louis MO
16. Université Grenoble Alpes, CNRS, IPAG, Grenoble France
17. Johns Hopkins University, Baltimore MD
18. George Mason University, Fairfax VA
19. University of California, Santa Cruz CA

# Abstract

Pluto's atmospheric haze settles out rapidly compared with geological timescales. It needs to be accounted for as a surface material, distinct from Pluto's icy bedrock and from the volatile ices that migrate via sublimation and condensation on seasonal timescales. This paper explores how a steady supply of atmospheric haze might affect three distinct provinces on Pluto. We pose the question of why they each look so different from one another if the same haze material is settling out onto all of them. Cthulhu is a more ancient region with comparatively little present-day geological activity, where the haze appears to simply accumulate over time. Sputnik Planitia is a very active region where glacial convection, as well as sublimation and condensation rapidly refresh the surface, hiding recently deposited haze from view. Lowell Regio is a region of intermediate age featuring very distinct coloration from the rest of Pluto. Using a simple model haze particle as a colorant, we are not able to match the colors in both Lowell Regio and Cthulhu. To account for their distinct colors, we propose that after arrival at Pluto's surface, haze particles may be less inert than might be supposed from the low surface temperatures. They must either interact with local materials and environments to produce distinct products in different regions, or else the supply of haze must be non-uniform in time and/or location, such that different products are delivered to different places.

# 1. Introduction

The discovery of extensive haze in Pluto's atmosphere was one of numerous striking findings from NASA's New Horizons Pluto encounter (Stern et al. 2015). The ultraviolet solar occultation data showed haze at altitudes up to at least 350 km above Pluto's surface (L.A. Young et al. 2018) and it can be seen in high phase visible wavelength images up to well over 200 km altitude (Gladstone et al. 2016; Cheng et al. 2017). The images reveal it to be divided into as many as 20 distinct layers, possibly due to the influence of orographic gravity waves. The haze is thought to have a relatively short residence time in Pluto's thin atmosphere before it settles to the surface, where it can be expected to accumulate. This paper explores the potential influence of haze particles on three very distinct terrain types on Pluto's encounter hemisphere, where the highest spatial resolution images and compositional maps were obtained (see Fig. 1). Cthulhu[1] is the largest of a group of dark, red patches that form a discontinuous belt along Pluto's equator. Portions of Cthulhu are heavily cratered and ancient, and it is mostly free of the seasonally mobile volatile ices that cover most of Pluto's surface. It thus offers a view of a thick, long-term accumulation of haze particles. In contrast, Sputnik Planitia is a smooth, bright, low-lying plain where convective overturn of volatile ice along with diurnal and seasonal cycles of sublimation and condensation refresh the surface on geologically short timescales. Lowell Regio is Pluto's northern polar region, where thick deposits rich in methane ice (Grundy et al. 2016a;

---

1 Names of Pluto system regions and features mentioned in this paper include a mix of official and informal names.

Howard et al. 2017) appear to have accumulated over intermediate timescales, possibly associated with Pluto's three million year Milankovitch-like seasonal cycles (Dobrovolskis et al. 1997; Earle & Binzel 2015; Hamilton et al. 2016; Earle et al. 2017; Stern et al. 2017).

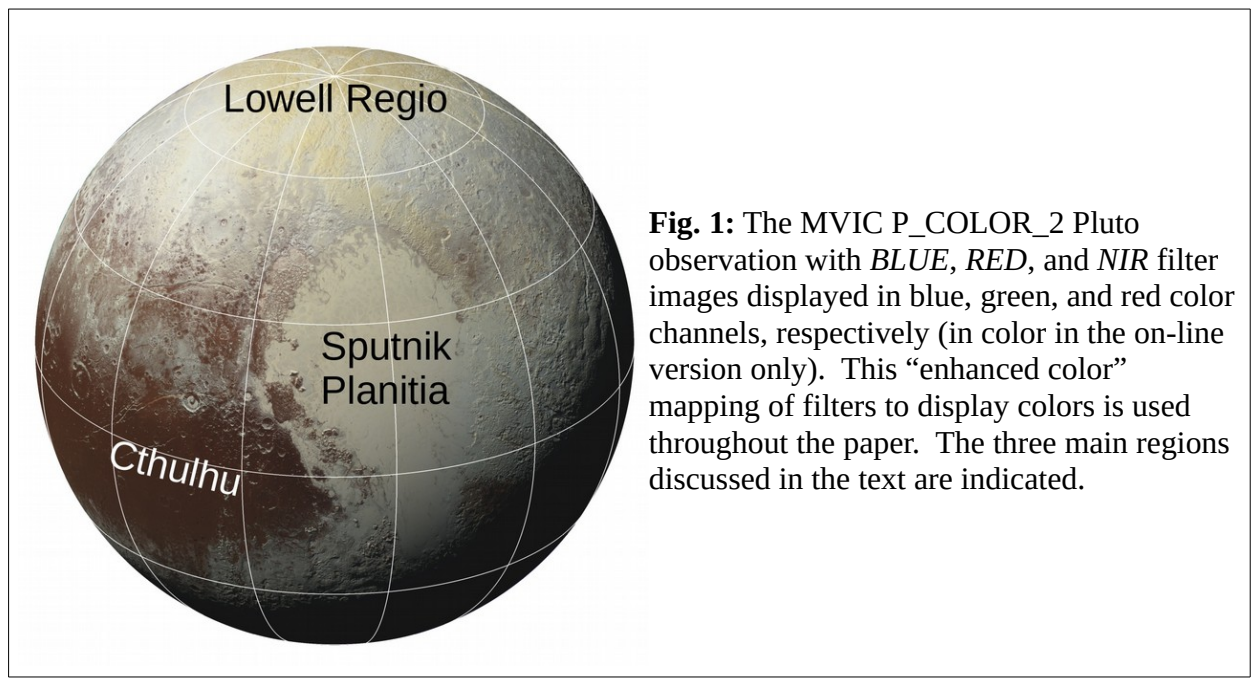

**Fig. 1:** The MVIC P_COLOR_2 Pluto observation with *BLUE*, *RED*, and *NIR* filter images displayed in blue, green, and red color channels, respectively (in color in the on-line version only). This "enhanced color" mapping of filters to display colors is used throughout the paper. The three main regions discussed in the text are indicated.

## 2. Haze Production and Composition

Pluto's haze originates in photolysis and radiolysis of gases in the upper atmosphere: chiefly $CH_4$, $N_2$, and CO. $N_2$ is the dominant atmospheric species near the surface, but $CH_4$ becomes increasingly abundant with altitude, matching the abundance of $N_2$ at an altitude of ~1400 km (Gladstone et al. 2016; L.A. Young et al. 2018). Methane is the primary chemical feedstock for haze production and is also the dominant molecular species escaping to space. Various energetic radiation sources produce radicals and ions from Pluto's atmospheric molecules, leading to creation of new compounds and eventually haze particles.

Solar ultraviolet light drives photochemistry, with Ly α photons (1216 Å, 10.2 eV) being an especially important source. The solar Ly α flux at Pluto's 39.5 AU mean heliocentric distance is ~$3 \times 10^{12}$ photons m$^{-2}$ s$^{-1}$ (Gladstone et al. 2015). Ly α photons readily break C-H bonds in methane through various pathways (to $CH_3(X) + H$, 5.7 eV; to $CH_2(a) + H_2$, 5.2 eV; and to $CH(X) + H_2 + H$, 1.14 eV) and are thus strongly absorbed by $CH_4$ in Pluto's upper atmosphere. In addition to direct solar illumination of Pluto's day side, Ly α photons are resonantly scattered by neutral H in the interplanetary medium, leading to diffuse irradiation of Pluto's night side by scattered Ly α comparable to the flux received on the day side (Gladstone et al. 2016; L.A. Young et al. 2018). Over Myr and longer timescales this resonantly scattered flux can be expected to vary as the Sun passes through different galactic environments. Ly α photons are not able to photodissociate the strong triple bonds N≡N in $N_2$ or C≡O in CO (9.8 and 11.1 eV, respectively). In the case of $N_2$, this is because there are no dipole-allowed transitions to repulsive states below the ionization threshold (15.6 eV). But shorter wavelength UV photons can break up $N_2$ through ionization and subsequent ion-molecule reactions, although their fluxes are lower. Higher energy x-ray photons (wavelengths < 1 nm) are scarcer still (Lisse et al. 2017), and highly variable with solar activity. They are more penetrating, and so contribute relatively little to atmospheric chemistry except near the surface, with the higher energy x-rays (< 0.1 nm) affecting Pluto's surface.

Another contributor to atmospheric chemistry is the plasma of protons and electrons streaming radially away from the Sun. Typical solar wind energies are in the keV range, though the distribution also includes higher energy particles (e.g., Cooper et al. 2003; Mewaldt et al. 2007). At Pluto's heliocentric distance during the 2015 New Horizons encounter, the flux of solar wind particles was highly time-variable, on the order of $10^9$ to $10^{10}$ particles $m^{-2} s^{-1}$ (Bagenal et al. 2016). Details of the plasma interaction with Pluto's atmosphere are uncertain, and much of it is diverted around Pluto due to the existence of a highly-conducting day-side ionosphere (Cravens & Strobel 2015), but it could be an important additional driver of radiolytic chemistry, with each keV particle delivering ~100 × the energy of a Ly α photon.

Cosmic rays provide yet another source of energetic particles, impinging on Pluto's atmosphere isotropically, with energies in the MeV through GeV range being especially important (e.g., Padovani et al. 2009). When one of these highly energetic particles collides with a molecule in the atmosphere (or on the surface, since the atmosphere is not thick enough to stop the more energetic particles) it triggers a cascade of lower energy secondary and tertiary particles that penetrate further, creating a substantial swath of damage: broken chemical bonds and excited radicals and ions that react to create new chemical species. At lower energies, this cascade is dominated by the component nuclei and electrons of the target atoms (elastic collisions), while at higher energies, atomic nuclei themselves can be disrupted, producing a shower of more exotic particles (inelastic collisions) (e.g., Johnson 1990). Estimates of cosmic ray penetration into the heliosphere suggest that Pluto currently receives relatively low fluxes of 100 keV to 10 MeV protons compared with objects orbiting closer to the heliopause (Cooper et al. 2003), but above ~100 MeV, the cosmic ray flux is less diminished by the heliosphere. Although cosmic ray fluxes are relatively low, the energies delivered by individual particles can be enormous, and over time they contribute to chemistry, too. Their fluxes also vary considerably as the Sun moves through different galactic environments.

Radicals, ions, and free electrons in Pluto's upper atmosphere/ionosphere interact to build heavier molecules, in much the same way as occurs on Titan, helped by attraction between oppositely charged ions (Quirico et al. 2008; Vuitton et al. 2009a; Lavvas et al. 2013). Macro-molecules as large as thousands of Da (atomic mass units), corresponding to sizes in the tens of nanometer range, are thought to form before they coagulate and settle out (Cheng et al. 2017).

Following Khare et al. (1984), we use the term *tholin* to refer to this complex, organic, macromolecular material.

The greater brightness of Pluto's haze at blue wavelengths is consistent with Rayleigh scattering by very small particles, but the strongly forward scattering phase function implies larger particles. These observational constraints can be reconciled if the tens of nm monomers join up to form micron-sized aggregates (e.g., Gladstone et al. 2016; Cheng et al. 2017). Particle settling times depend strongly on size. For the 1 µm size range favored by observations of forward scattering, they are relatively short, in the range of Earth days to months (Cheng et al. 2017; Gao et al. 2017; Zhang et al. 2017), though it may take of order a few Earth years for them to grow to that size before they settle out.

At altitudes above a few hundred km, New Horizons found Pluto's atmosphere to be ~70 K, cooler than had been expected (e.g., Gladstone et al. 2016; Hinson et al. 2017). The dominant cooling mechanism appears to be thermal infrared radiation from the haze particles themselves (Zhang et al. 2017). Below about 700 km altitude, frequent collisions with gas molecules keep haze particles thermally coupled to the atmosphere.

Additional cooling comes from emission from gaseous hydrogen cyanide (HCN), acetylene ($C_2H_2$), water ($H_2O$), and possibly other species (Gladstone et al. 2016; Strobel and Zhu 2017). Although these are not thermodynamically stable as gases at such low temperatures, they could be resupplied through photochemistry and/or meteoritic infall. They go on to react with other species or stick to the growing haze particles through condensation or more likely adsorption (e.g., Gao et al. 2017; Luspay-Kuti et al. 2017; Strobel and Zhu 2017). The details are highly uncertain as to how these molecules are incorporated into the haze particles and whether they retain their identities as distinct molecules or are chemically bonded into the macro-molecular tholin. From photochemical models, Wong et al. (2017) estimated the present-day production of species contributing to the growth of the haze particles, finding them to be dominated by $C_2$ hydrocarbons acetylene ($C_2H_2$), ethylene ($C_2H_4$), and ethane ($C_2H_6$), along with a variety of other hydrocarbons and nitriles (see Table 1). Since the Wong et al. model neglected ionospheric chemistry, these production rates may be underestimates.

**Table 1**

| Species name and formula | | Production (g cm$^{-2}$ Gyr$^{-1}$) |
|---|---|---|
| Acetylene | $C_2H_2$ | 179 |
| Ethylene | $C_2H_4$ | 95 |
| Ethane | $C_2H_6$ | 62 |
| Propyne | $CH_3C_2H$ | 48 |
| Hydrogen cyanide | HCN | 35 |
| Benzene | $C_6H_6$ | 34 |
| Diacetylene | $C_4H_2$ | 26 |
| Propene | $C_3H_6$ | 8 |
| Methyl-cyanoacetylene | $CH_3C_3N$ | 6 |
| Cyanoacetylene | $HC_3N$ | 4 |

Table adapted from Wong et al. (2017).

Lower down, the atmospheric temperature reaches a maximum of ~110 K around 30 km altitude (Hinson et al. 2017). Some evolution as haze particles settle through this warm zone seems inevitable. The temperature is high enough to cause sublimation of more volatile hydrocarbons and nitriles that condensed onto the haze particles higher up in the atmosphere, but only if they condensed rather than being more strongly adsorbed or even chemically bonded into the macromolecular tholin (e.g., Luspay-Kuti et al. 2017). Sublimation loss would shrink the haze particles and slow their settling, as well as modifying their compositions by distilling off the more volatile constituents. 110 K also exceeds the zero pressure melting points of some of the constituent hydrocarbon species, specifically $C_2H_4$ (104 K), $C_2H_6$ (90 K), and $C_3H_6$ (88 K). Unmelted species may partly dissolve into those that do melt, and melting point depression can be expected in such mixtures (for instance, the $CH_4$-$C_2H_6$ eutectic melting point is 18 K below the melting points of the pure species; Hanley et al. (2017)). Partial melting could allow particles to evolve from fractal aggregate shapes into more compacted spherical forms, due to surface tension, with implications for their scattering properties and settling rates. Warmed haze particles might also be more sticky and could coalesce into even larger aggregates if they contact one another.

In the last few km above the planet's surface, the atmospheric temperature declines quickly toward ~40 K, approaching vapor pressure equilibrium with $N_2$ ice on the surface. Haze particles descending through this much colder near-surface region will refreeze any partial melting and return to accreting gaseous hydrocarbon and nitrile molecules, growing larger again, and falling out even more quickly. Cheng et al. (2017) report altitude-dependent photometric properties for the haze in the lower atmosphere that are likely related to the abruptly changing thermal environment the settling haze particles experience on their descent through the last few tens of km above the surface.

Summing up the present-day production rates in Table 1, we obtain ~500 g cm$^{-2}$ Gyr$^{-1}$, all of which is presumed to accumulate on Pluto's surface. Those numbers were for production rates during the 2015 encounter, when Pluto was 32.9 AU from the Sun. Production should scale with the UV flux, which falls off approximately as the square of distance from the Sun, so a seasonal average of ~350 g cm$^{-2}$ Gyr$^{-1}$ can be obtained by scaling by the square of the encounter distance divided by the square of Pluto's mean distance from the sun of 39.5 AU. Assuming an average density of 1 g cm$^{-3}$ and neglecting porosity, this material would coat the surface to a thickness of 3.5 m in 1 Gyr, or about 14 m over the age of the solar system. A one-micron thick monolayer of haze particles would take about 3 Earth centuries to accumulate, roughly a Pluto year.

It is unclear how time-variable the haze production would have been over Pluto's history. The UV flux at Pluto varies from perihelion to aphelion and with the solar cycle, and the geometric cross section and $CH_4$ content of Pluto's atmosphere likely also varies seasonally. So this estimate of ~14 m of haze produced over the age of the solar system must be regarded as highly uncertain. This tally also omits the macromolecular tholin component of the haze particles, unless all of the light hydrocarbons and nitriles are consumed in production of tholin. It is unknown what fraction of Pluto's haze particles would be composed of tholin. In Titan haze, Lavvas et al. (2013) estimate it to be 7 to 10%. Critical to formation of tholin in haze is the presence of $CH_4$ at altitudes where ion chemistry occurs, ensuring the formation of the nitriles and negative nitrile ions that are precursors to tholin formation. This condition holds in Pluto's current upper atmosphere, unlike in Triton's (Strobel & Zhu 2017), but Pluto's atmosphere might not always be suitable for tholin production.

Small amounts of additional material arrive at Pluto's surface in the form of dust particles from the Kuiper Belt (Poppe 2015), and ejecta debris from the small satellites (Porter & Grundy 2015). The composition of this material is probably dominated by $H_2O$ ice, though some carbonaceous and silicate material is likely also present. These sources are estimated to contribute of order 5 mm and 10 mm of dust over the age of the solar system, for Kuiper belt and small satellite sources, respectively (Poppe 2015; Porter & Grundy 2015). These quantities are very small compared with a ~14 m accumulation of haze particles, but they could still be important contributors to haze production as sources of oxygen and in providing potential nucleation sites in the upper atmosphere.

## 3. Model Haze Particle

To simulate potential effects of haze particles on the observable characteristics of Pluto's various surface regions using multiple-scattering radiative transfer models, we would need to know the wavelength-dependent single scattering albedo $w$ and single scattering phase function $P(g)$ of the haze particles, which in turn depend on their size, composition, shape, and internal configuration. Considering the numerous uncertainties associated with their formation, composition, and subsequent modification on descent through Pluto's atmosphere, no definitive model is possible. But a crude model haze particle can be constructed for use in radiative transfer models by considering that we expect the ingredients to include some amount of macromolecular tholin as well as some lighter hydrocarbons and nitriles, and that we expect these ingredients to be mixed on spatial scales in the nanometer to micrometer size range. At visible wavelengths, light hydrocarbon and nitrile ices are generally transparent and colorless, while unsaturated linkages in tholins absorb strongly at blue wavelengths (e.g., Rao et al. 1975; Imanaka et al. 2004; Quirico et al. 2008). The tholin component should thus dominate visible wavelength spectral behavior. We can treat the haze particles as being effectively a combination of just 2 ingredients: a pigment plus a transparent matrix. Fig. 2 illustrates some potential configurations. For materials mixed at spatial scales below the wavelength of light, effective optical constants of the aggregate mixture can be approximated using effective medium theory (e.g., Garnett 1904). For the transparent ice, we assume $n = 1.4$ and $k = 0$ for the real and imaginary parts of its refractive index, respectively, over the visible to near-infrared wavelengths observed by New Horizons. For the tholin, plausible candidates would be the Titan analog tholin optical constants measured by several groups (e.g., Khare et al. 1984; Ramirez et al. 2002; Imanaka et al. 2004; Vuitton et al. 2009b; Sciamma-O'Brien et al. 2012). These materials are laboratory analogs for macro-molecular material produced in Titan's upper atmosphere under conditions similar to those in Pluto's upper atmosphere (e.g., Cheng et al. 2017). The mixing ratio of tholin to ices in Pluto's haze particles is unknown and thus is left as a free parameter in our model haze particle.

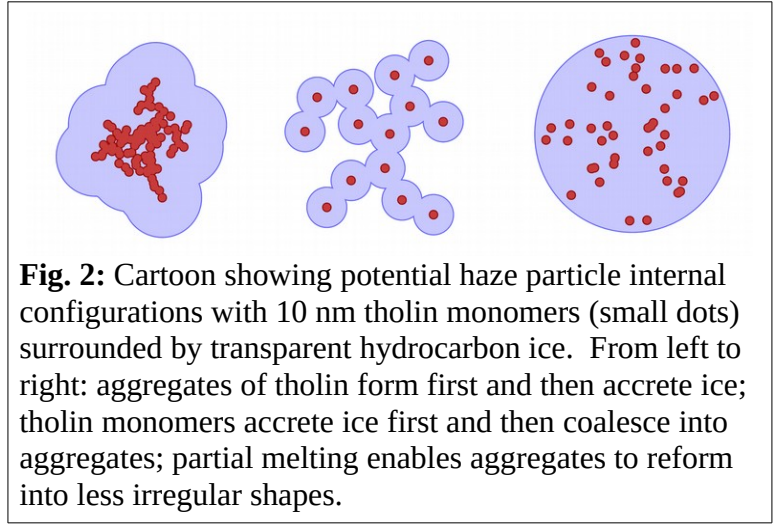

**Fig. 2:** Cartoon showing potential haze particle internal configurations with 10 nm tholin monomers (small dots) surrounded by transparent hydrocarbon ice. From left to right: aggregates of tholin form first and then accrete ice; tholin monomers accrete ice first and then coalesce into aggregates; partial melting enables aggregates to reform into less irregular shapes.

## 4. Data

This paper focuses on Pluto's encounter hemisphere, where relatively high spatial resolution data sets were obtained during the last few hours prior to New Horizons' closest approach to Pluto on 2015 July 14. We primarily consider data from the Ralph instrument, which consists of two separate focal planes fed by a single 75 mm aperture $f$/8.7 telescope (Reuter et al. 2008). Visible and near-infrared light out to 1.1 µm is reflected by a dichroic beamsplitter to the Multispectral Visible Imaging Camera (MVIC). Longer infrared wavelengths are transmitted to the Linear Etalon Imaging Spectral Array (LEISA).

MVIC is Ralph's visible and near infrared imager. Its focal plane has seven CCD arrays, six of which are 5024 × 32 pixel devices operated in Time Delay Integration (TDI) mode. In this mode, MVIC's field of view is swept across the scene, perpendicular to the long axis of the CCDs, and the charges are read out at rates matched to the scan motion. The result is an image strip 5024 pixels wide and arbitrarily long. Two of the TDI CCDs are panchromatic, covering wavelengths from 400 to 975 nm. Four have color filters affixed to them and are operated simultaneously as a group, enabling 4-color images to be recorded. The filter names and wavelengths are "*BLUE*" 400-550 nm, "*RED*" 540-700 nm, "*NIR*" (near-infrared) 780-975 nm, and "*CH4*" (methane) 860-910 nm.

The primary MVIC data set used in this paper is P_COLOR_2 (Fig. 1). This observation was obtained 2015 July 14 at 11:10:52 UT (observation mid-time), when the spacecraft was at a range of 34,000 km from Pluto and the mean phase angle was 39°. The image scale was 650 m pixel$^{-1}$. This dataset's unique Mission Elapsed Time (MET) identifier is 0299178092. The four MVIC filter images were processed separately through the standard New Horizons Science Operations Center (SOC) pipeline for flat fielding, and then radiometrically calibrated to specific intensity (I/F; Chandresekhar 1960) and destriped as described by Howett et al. (2017) and Olkin et al. (2017). MVIC's *RED*, *NIR*, and *CH4* filter images were geometrically registered to the contemporaneous *BLUE* image using version 3.5.0.7383 of the Integrated Software for Images and Spectrometers (ISIS3) package from the United States Geological Survey, and the geometry of the *BLUE* image adopted for subsequent analysis. The four color CCDs are adjacent to one another on the focal plane, so they do not view the same point on the target exactly simultaneously. Owing to spacecraft and target motion during the scan, each CCD sees slightly different geometry. Neglecting those differences introduces little error, however, since at the scan rate of 1045 μrad s$^{-1}$, all four CCDs are swept across a point in the scene in just six seconds.

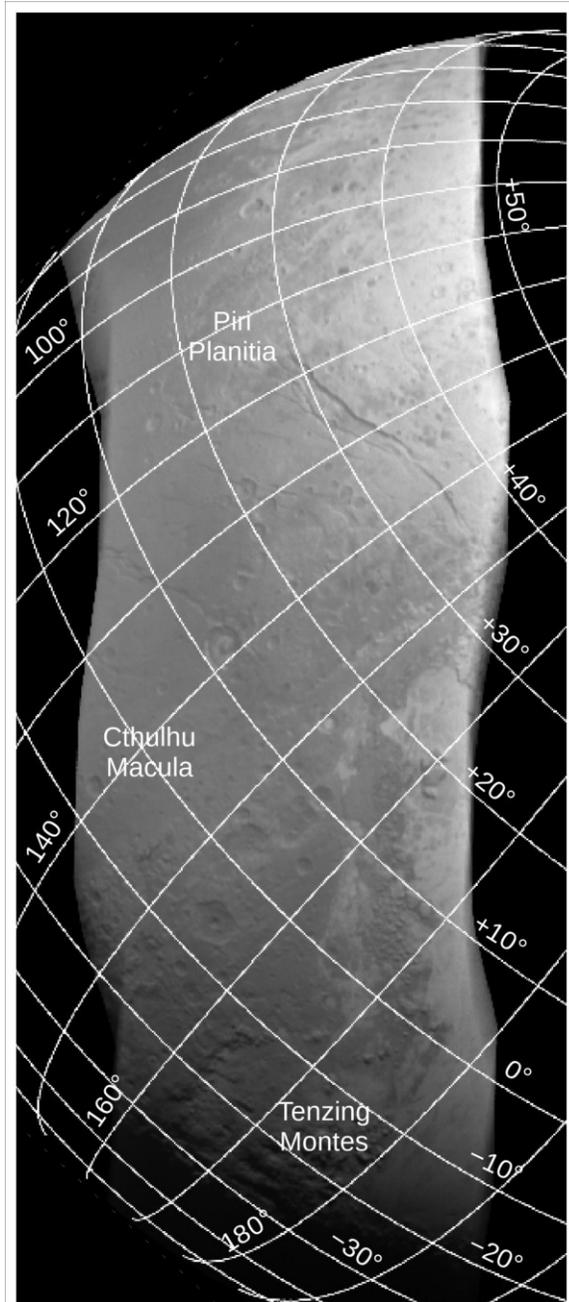

LEISA is Ralph's infrared imaging spectrometer. It has a 256 × 256 HgCdTe detector array with a pair of linear variable interference filters mounted directly onto it such that each row of the detector is sensitive to a different wavelength of infrared light. The primary filter covers wavelengths from 1.25 to 2.5 μm spanning about 200 rows of the array. The resulting spectral resolution (λ/Δλ) is about 240. A higher spectral resolution filter covering wavelengths from 2.1 to 2.25 μm occupies the remaining rows, but data from that portion of the array have proved more difficult to calibrate and were not used in this paper. In order to build up a spectral image cube, the detector array is swept across the target scene while a series of frames is recorded. The frame rate and scan rate are coordinated such that the scene shifts by about a pixel from one frame to the next.

For LEISA, the highest spatial resolution Pluto observation is P_LEISA_HIRES, obtained 2015 July 14 at 10:56:19 UT, from a mean range of 45,000 km and a mean phase angle of 33°. The mean image scale was 2.7 km pixel$^{-1}$. The dataset's MET unique identifier is 0299176809. At the mid-time of the scan, LEISA's 256 pixel field of view covered ~690 km at the distance of Pluto, so only a strip of the encounter hemisphere was observed, extending toward the south and east from the lit limb, over Piri Planitia, eastern Cthulhu, and Tenzing Montes to the terminator, as illustrated in Fig. 3.  The standard SOC pipeline (e.g., Peterson et al. 2013; Howett et al. 2017; Olkin et al. 2017) provides radiometrically calibrated data points, each with a distinct wavelength and spatial footprint. These points can be projected onto a map using ISIS3.  We projected them to the point-perspective view from the spacecraft's location at the mid-time of the scan, with 1 km sample spacing at image center, and adopted the mid-scan geometry.  This is a worse approximation for LEISA than it is for MVIC since LEISA scans take much longer to complete.  The P_LEISA_HIRES scan was done a factor of ten slower, at 105 µrad s$^{-1}$.  The entire scan lasted 16 minutes, of which at least some portion of Pluto was within the field of view for almost 10 minutes.  Latitude and longitude angles are unaffected, since ISIS3 projects each pixel to its correct location in latitude and longitude (assuming a spherical Pluto with radius 1,188.3 km; Nimmo et al. 2017).  But incidence angles change slightly over the time of the scan due to Pluto's rotation.  Pluto rotates through 0.39° of longitude in 10 minutes.  Emission and phase angles are even more affected, since they depend on the spacecraft's location.  Over the portion of the scan when Pluto was being observed, the spacecraft traveled 7,900 km relative to Pluto. During that time, the phase angle changed by 3.2°, and the sub-spacecraft point's latitude and longitude shifted by 2.4° and 2.0°, respectively.  The effect of assuming fixed mid-scan geometry instead of time-varying geometry over the course of the scan is to tacitly accept emission and phase angle errors on the scale of a few degrees.

**Fig. 3:** The P_LEISA_HIRES observation, showing the median of all LEISA wavelengths.

We also use data from one other LEISA scan that has not previously been published, P_MULTI_DEP_LONG_1.  This observation was done about four hours after closest approach, looking back at the night side of Pluto at a mean phase angle of 169°.  The mid-scan time was 2015 July 14 15:27:22 UT, corresponding to a 181,000 km range and an image scale of 11 km pixel$^{-1}$.  The MET unique identifier is 0299193339.

## 5. Three Provinces

Pluto's encounter hemisphere features a number of very distinct regions, with diverse compositions, morphologies, and ages (see Fig. 1).  We consider three of the most prominent provinces here, exploring how each might be influenced by fallout and accumulation of haze particles.

In all cases, we assume constant haze production at rates scaled from present-day estimates, and that surface deposition is spatially uniform, at least when averaged over time.  These uniformitarian assumptions are likely oversimplifications, but perhaps not egregiously so.  To the extent that haze production is limited by the flux of solar Ly α, it might not vary all that much over time.  Of course, if there were past epochs when Pluto's atmosphere collapsed, or when its CH$_4$ content was negligible, removing the main source of Ly α opacity and also the main feedstock for photochemistry, haze production could have ceased or been greatly diminished and the energetic radiation no longer blocked by the atmosphere would act directly on Pluto's surface ices (e.g., Kim & Kaiser 2012; Wu et al. 2012; Materese et al. 2014). With settling rates being slow compared with wind speeds, haze deposition should also be relatively evenly distributed, though local factors could complicate this picture, such as the diurnal cycles of sublimation/condensation and katabatic winds predicted by global circulation models (e.g., Bertrand & Forget 2017; Forget et al. 2017).

## 5.1 Cthulhu

Pluto's equatorial latitudes feature a prominent belt of large, dark, reddish maculae (Latin for spots or stains). The largest of these is Cthulhu, which extends more than a third of the way around the planet, from longitude 20° E through 160° E. Cthulhu is mostly depleted in volatile ices, apart from a few bright, possibly seasonal deposits localized on mountain peaks and crater rims (e.g., Grundy et al. 2016a). Portions of Cthulhu are densely cratered, implying an ancient surface (Moore et al. 2016; Robbins et al. 2017), though many areas, especially towards the core of Cthulhu, look smooth and uniformly dark at spatial scales below a few tens of km. Towards Cthulhu's periphery, the ubiquetous dark material appears to become less concentrated and $H_2O$ ice is increasingly detectable from its absorptions at 1.5 and 2 µm. The $H_2O$ ice could be Pluto's bedrock, revealed by the dark material becoming patchy and/or thinner toward the periphery of Cthulhu, perhaps due to erosive action of seasonal volatile ice deposits. Or the $H_2O$ ice could be superposed on top of the dark material, though it is not clear what mechanism would do that just along the periphery. There is no obvious morphological or topographic contrast between Cthulhu and adjacent terrains (Schenk et al. 2018). These aspects are suggestive of a superficial coating of highly pigmented material in Cthulhu, rather than a distinct geological stratum being exposed there (e.g., Sekine et al. 2017).

From the estimated haze production rates discussed earlier, ~3.5 kg settles onto each square meter of otherwise inert surface in 1 Myr. This would produce a ~3.5 mm thick coating in 1 Myr, and over 4 Gyr, a ~14 m thick blanket of haze particles would accumulate. A variety of processes could affect the accumulating haze particles after their arrival at the surface. In the remainder of this section, we consider thermal processes such as sublimation and sintering, effects of energetic radiation, and the apparent absence of craters that expose substrate material. We also examine the photometric and spectral characteristics of Cthulhu's surface.

Low albedo, volatile-free regions on Pluto's equator may get relatively warm, especially if they have low thermal inertias. An upper limit can be estimated by assuming negligible thermal inertia such that a region is in instantaneous thermal equilibrium between insolation and emission. When Pluto's equinox coincides with its perihelion (as it does at present; Earle & Binzel 2015), an equatorial region with a bolometric bond albedo of 0.1 (Buratti et al. 2017) and emissivity assumed to be 0.9 can be calculated to reach 72 K at perihelion. Vapor pressures at that temperature for potential haze constituents $C_2H_4$ and $C_2H_6$ are 0.0009 and 0.0046 Pa, respectively (Fray & Schmitt 2009). Comparing these non-negligible vapor pressures with Pluto's present-day surface pressure of ~1 Pa suggests that the haze particles on the ground in Cthulhu could evolve via sublimation loss of their more volatile constituents, potentially leading to greater concentration of tholins over time. Another possibility is sintering (e.g., Eluszkiewicz et al. 1998), considering that 72 K is not far below the zero pressure melting points of pure simple hydrocarbons like $C_2H_6$ (90 K), $C_3H_8$ (85 K), and $C_2H_4$ (104 K), and that partial melting of hydrocarbon mixtures can occur at even lower temperatures (Hanley et al. 2017). Sintering could bind the deposited material into a more durable surface and reduce the possibility of aeolian transport (e.g., Telfer et al. 2018). But it would also increase the thermal conductivity and thus the thermal inertia, reducing the maximum temperature reached, so it could be a self-limiting process. It is worth noting that Buratti et al. (2017) estimated Cthulhu's bond albedo as 0.1 from visible wavelength LORRI images (350-850 nm; Cheng et al. 2008), but considering Cthulhu's steep rise in albedo toward longer wavelengths, that figure should probably be taken as a lower limit for the bolometric bond albedo. A higher value would translate to lower maximum temperatures reached.

Unlike Ly α photons and lower energy solar wind particles, cosmic rays with energies above ~30 MeV can penetrate through Pluto's present-day atmosphere and drive chemical evolution on the surface. X-rays could also contribute to processing of surface material, though their fluxes are low (Lisse et al. 2017). Estimates of cosmic ray radiolysis at Pluto (Strazzulla et al. 1984) find that depths down to ~5 mm will receive chemically significant doses in 100 Myr, during which time a much greater ~350 mm of new haze particles would be deposited. A comparable estimate can be arrived at by integrating the flux at 40 AU in protons over the energy range from 30 MeV up through 10 GeV from Cooper et al. (2003; Fig.3A). This totals ~2 GeV cm$^{-2}$ s$^{-1}$, which will affect chemistry in a layer of ice many cm thick. The degree of processing reached prior to burial depends on how this energy is distributed versus depth, but simplistically assuming it to be distributed uniformly over the uppermost 10 cm, we find that these cosmic rays deliver around 4% of the radiation needed to process the material to the 6 eV Da$^{-1}$ dose considered to be chemically significant by Cooper et al. (2003) during the ~30 Myr that it takes for another 10 cm of haze to accumulate. From these estimates, it appears that while cosmic rays can enable some further processing of Cthulhu's accumulating haze, that processing does not progress particularly far before the products are buried by the arrival of additional fresh haze particles. The implication is that there must already be appreciable tholin present in the haze by the time it arrives at the surface, since there does not appear to be enough radiation reaching the surface to create much of it from the light hydrocarbon component of Pluto's haze, unless there are episodes when the atmosphere fails to screen out UV. Astrophysical events such as nearby supernovae or the Sun's passage through a region of dense galactic gas that compresses the heliosheath could also play a role in increasing energetic radiation at Pluto's surface, but with new haze accumulating as fast as we estimate that it does, there would have to have been a recent event to have modified the optically active uppermost stratum of Cthulhu observed by New Horizons.

On Iapetus, small, fresh craters with bright ejecta were seen in high resolution (~10 m pixel$^{-1}$) images of the dark exogenous material that coats the equatorial and mid-latitude regions of its leading side (Denk et al., 2009). The small, bright craters were used to determine that the dark material was a relatively thin coating. The fact that no fresh craters are seen in Cthulhu exposing bright ice from below probably attests to the darkening being a relatively rapid and ongoing process, consistent with accumulating dark haze particles. The lack of such craters could also mean that the dark material is relatively thick, but high resolution images comparable to those for Iapetus are not available for Cthulhu. A few small craters do expose bright substrate material in Mordor Macula, Charon's dark red northern polar deposit (Grundy et al. 2016b), indicating that the accumulation rate of dark material in Cthulhu is likely much more rapid than it is in Mordor. That observation is consistent with our estimate of ~14 m accumulation over 4 Gyr in Cthulhu versus just ~0.3 m in Mordor (Grundy et al. 2016b).

One quantitative technique to understand the origin of the low albedo material on Cthulhu is to derive the macroscopic roughness of the surface. Icy and rocky surfaces with little infall of exogenously-produced material tend to have rough surfaces, with mean slope angles of ~30° (Buratti 1985; Helfenstein and Veverka 1987; Verbiscer and Veverka 1992; Domingue et al. 1995). In contrast, the low-albedo hemisphere of Iapetus, which is coated by dust from the Phoebe ring (Buratti and Mosher 1995; Verbiscer et al. 2009; Tamayo et al. 2011) is smooth, with a mean slope angle of only 6°, presumably due to infilling of rough facets by exogenic dust (Lee et al. 2010). Similarly, an analysis of high and low albedo terrains on Titan showed that the low albedo terrain was only about half as rough as the high albedo terrain, presumably due to deposits of organic particles formed in Titan's haze layers (Buratti et al. 2006).

To model the roughness of Cthulhu, we employed a crater roughness model (Buratti and Veverka 1985), which is similar to the more widely used mean slope model of Hapke (1984). The former model places paraboloidal craters defined by a depth-to-radius ratio on the surface: it can easily be compared to the Hapke roughness model by taking the first derivative of the equation that defines the shape of the crater and equating it to a mean slope angle. Scans of I/F were extracted from Cthulhu with associated values of the solar phase angle and radiance incident and emission angles. The scans were fit to the model by a least-squares method to derive the best depth-to-radius ratio, as shown in Fig. 4. The best-fit ratio is 0.04 ± 0.01, corresponding to a mean slope angle of 2° ± 1°. This number is not only exceedingly low, but similar to the best-fit depth-to-radius value of of 0.084 (corresponding to a mean slope angle of 6°) for the low albedo hemisphere of Iapetus (Lee et al. 2010).

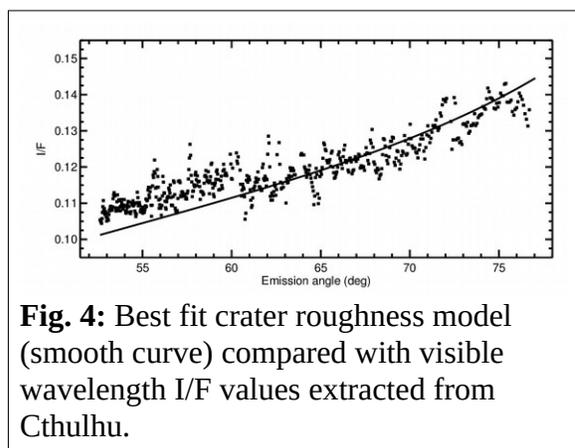

**Fig. 4:** Best fit crater roughness model (smooth curve) compared with visible wavelength I/F values extracted from Cthulhu.

The smoothness of Cthulhu is also attested to by its forward scattering at near infrared wavelengths as observed in the P_LEISA_HIPHASE observation at a mean phase angle of 169° (see Fig. 5). This observation shows enhanced brightness where the illuminated crescent slices across Cthulhu. It also shows that the blue coloration of Pluto's haze reported by Cheng et al. (2017) extends into near-infrared wavelengths, and that infrared scattering by haze particles is detectable up to as high as ~150 km altitude before being lost in the noise.

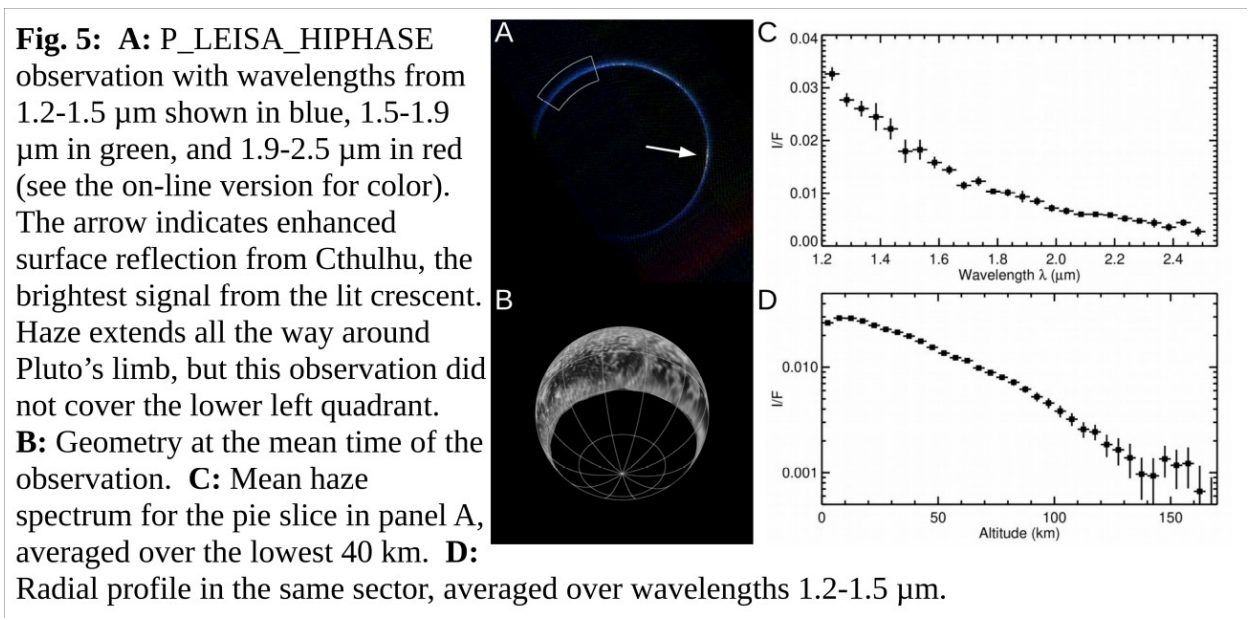

**Fig. 5: A:** P_LEISA_HIPHASE observation with wavelengths from 1.2-1.5 µm shown in blue, 1.5-1.9 µm in green, and 1.9-2.5 µm in red (see the on-line version for color). The arrow indicates enhanced surface reflection from Cthulhu, the brightest signal from the lit crescent. Haze extends all the way around Pluto's limb, but this observation did not cover the lower left quadrant. **B:** Geometry at the mean time of the observation. **C:** Mean haze spectrum for the pie slice in panel A, averaged over the lowest 40 km. **D:** Radial profile in the same sector, averaged over wavelengths 1.2-1.5 µm.

With its relatively low rates of radiolytic and thermal processing, at least during present-day climate and radiation conditions, Cthulhu may be an ideal place to study Pluto's haze particles. They appear to form a deep accumulation on this ancient surface that is relatively uncontaminated by the spectral signatures of Pluto's seasonally-mobile volatile ices or bedrock $H_2O$ ice. The appearance is generally dark and red, but it is not perfectly homogeneous. Fig. 6 shows a hard stretch on P_COLOR_2 in eastern Cthulhu. Slightly darker shades are evident in many crater floors and also following the courses of valley floors. This distribution of shading is suggestive of preferential accumulation of the reddish material in topographic lows, perhaps under the influence of winds (e.g., Telfer et al. 2018) or through the erosive action of seasonally deposited volatile ices. But the difference in color between smooth plains and darker, low-lying areas is subtle, as indicated by the color ratios in the right panel of Fig. 6.

As reported by Olkin et al. (2017), some of the reddest colors in Cthulhu occupy a core region extending approximately from 120° to 145° E longitude and from −8° to +4° in latitude. With lower resolution LEISA data, Schmitt et al. (2017) show a well-populated mixing trend between the spectral characteristics of the core of Cthulhu and more $H_2O$-rich compositions along its periphery. The LEISA data place the core region further to the east than the MVIC colors do, at ~160° E longitude. The longitudinal difference between maximum redness in MVIC images and maximum tholin spectral signature in LEISA data is not yet understood, but could perhaps be a photometric effect of the incidence and emission angles that vary across the scene. Both instruments show the material becoming progressively more diluted with substrate material towards the northern and southern boundaries of the macula, suggesting that the Cthulhu core region is the place to look for haze particles with the least contamination by other materials. ROI 5 was selected to represent that region. The colors there are quite similar to those of ROIs 1 and 2, corresponding to topographic lows toward the eastern edge of the reddest core zone based on MVIC colors.

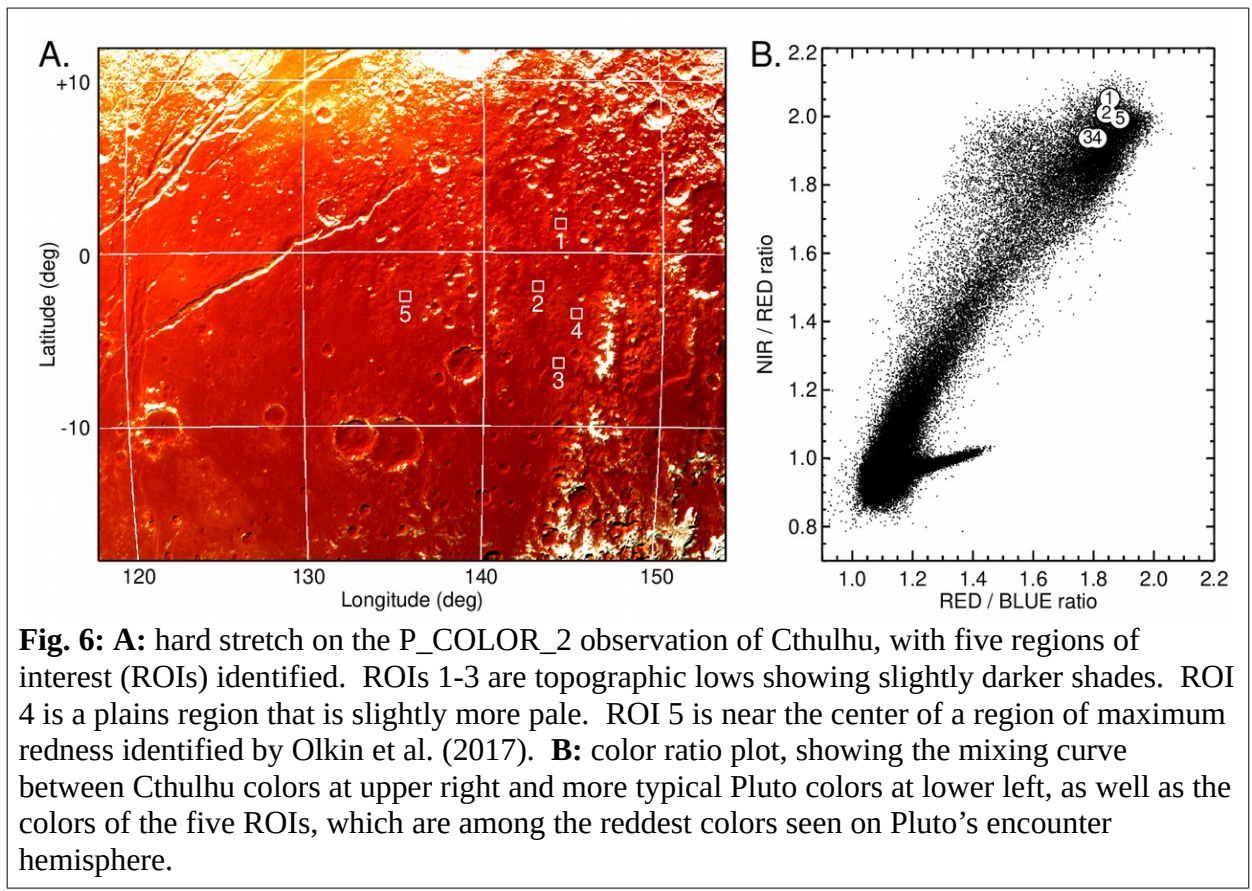

**Fig. 6: A:** hard stretch on the P_COLOR_2 observation of Cthulhu, with five regions of interest (ROIs) identified. ROIs 1-3 are topographic lows showing slightly darker shades. ROI 4 is a plains region that is slightly more pale. ROI 5 is near the center of a region of maximum redness identified by Olkin et al. (2017). **B:** color ratio plot, showing the mixing curve between Cthulhu colors at upper right and more typical Pluto colors at lower left, as well as the colors of the five ROIs, which are among the reddest colors seen on Pluto's encounter hemisphere.

Cook et al. (2018) analyzed lower spatial resolution LEISA Pluto spectra of the entire encounter hemisphere, specifically looking for signatures of non-volatile materials. They reported evidence for $H_2O$ ice absorption bands at 1.5 and 2 µm towards the periphery of Cthulhu, possibly material from Pluto's mantle mixed up into the reddish material through erosion by seasonally deposited volatile ices or some other geological process, or simply exposed in outcrops below the resolution limit of New Horizons' cameras. They also found evidence for absorption by $C_2H_6$ ice, especially in eastern Cthulhu, including the region where the Schmitt et al. (2017) LEISA compositional maps place the Cthulhu "core". Ethane was proposed as representative of all alkanes heavier than methane, but confidence in that interpretation is boosted by its identification in higher spectral resolution ground-based observations around the same longitude (Holler et al. 2014). Interestingly, Cook et al. also found tentative evidence for methanol ice ($CH_3OH$), also mostly in eastern Cthulhu. Wong et al. (2017) did not report expected abundances for oxygenated species among the photochemical constituents of haze particles, but they mention the possibility of oxygenated compounds, since oxygen is potentially available for photochemistry and ion-molecule chemistry, both from Pluto's atmospheric CO and from infalling $H_2O$-rich dust particles from the Kuiper belt and from the small satellites.

With its higher spatial resolution, the P_LEISA_HIRES observation has the potential to reveal details of regional variation in the materials that make up Cthulhu's reddish coating. But after excluding isolated patches of volatile ices, the LEISA spectra do not show a great deal of spectral variability within Cthulhu beyond the trends already reported by Schmitt et al. (2017) and Cook et al. (2018), consistent with the homogeneity seen in MVIC colors. Figure 7 compares spectra of the five ROIs, relative to a mean Cthulhu spectrum. ROI 1 might show a hint more of $H_2O$ absorption at 2 µm, while ROI 5 might show a hint more hydrocarbon absorption around 2.4 µm, but the differences are not especially striking.

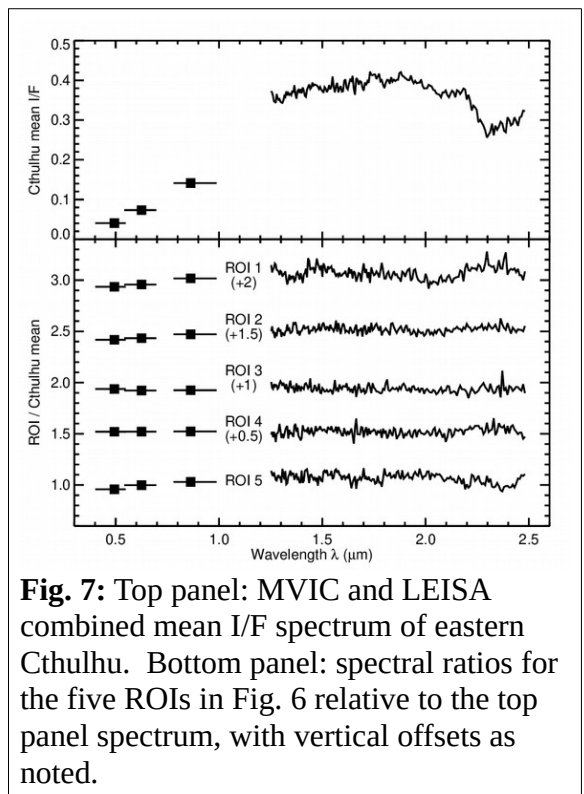

**Fig. 7:** Top panel: MVIC and LEISA combined mean I/F spectrum of eastern Cthulhu. Bottom panel: spectral ratios for the five ROIs in Fig. 6 relative to the top panel spectrum, with vertical offsets as noted.

We next consider whether Cthulhu's colors can be reproduced using our simple haze particle model. Using effective medium theory, we computed effective *n* and *k* for a range of tholin content from 1% through 90% relative to transparent hydrocarbon ice and then adjusted the particle size in a Hapke model (e.g., Hapke 1993) to get the right I/F range for the relevant illumination and observation geometry. Smaller particle sizes make the surface brighter while larger ones make it darker. We used grain sizes larger than the 1 µm generic haze particle size, since Hapke models assume the ray optics regime with grain sizes larger than the wavelength, and, as mentioned earlier, it is possible that haze particles could become sintered into larger aggregates. Even allowing this degree of freedom, the match to MVIC colors of these models was generally poor, being too dark in the *BLUE* filter and too bright in the *NIR* filter. This mismatch shows that the Titan tholin is too strongly pigmented for the assumed model configuration. The excess coloration could be remedied by introducing an additional neutral absorber, such as carbonaceous material. A material that absorbs more strongly at MVIC *NIR* wavelengths would also work. Although continuing radiolysis or photolysis of tholins is known to produce dark, neutral material similar to amorphous carbon, such a high degree of processing seems difficult to reconcile with the limited residence time of macromolecular material in Pluto's atmosphere and its rapid burial rate on the surface. Spectrally neutral, non-absorbing scatterers can similarly suppress color contrast, and are perhaps more plausible, since voids in an aggregate (such as the middle cartoon in Fig. 2) behave as neutral scatterers if they are large compared to the wavelength, $\lambda$, and like Rayleigh scatterers if they are much smaller. Neutral scattering can be introduced via Hapke's *s* parameter in the equivalent slab single particle model, with units of inverse length, and Rayleigh scattering by microscopic voids can be simulated by making *s* proportional to $\lambda^{-4}$. A simulation with 25% tholin haze in a matrix of transparent hydrocarbon ice, aggregated into 100 µm grains, and with $s = 10{,}000$ cm$^{-1}$ (for all wavelengths) matched the MVIC colors reasonably well, but is by no means a unique solution, and this and other models using Titan tholin all fail to match the high reflectance at LEISA wavelengths.

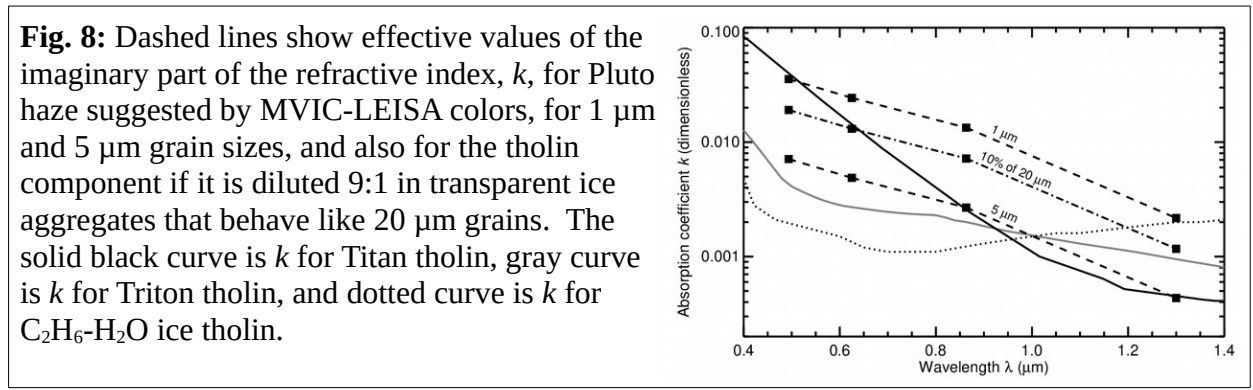

**Fig. 8:** Dashed lines show effective values of the imaginary part of the refractive index, *k*, for Pluto haze suggested by MVIC-LEISA colors, for 1 µm and 5 µm grain sizes, and also for the tholin component if it is diluted 9:1 in transparent ice aggregates that behave like 20 µm grains. The solid black curve is *k* for Titan tholin, gray curve is *k* for Triton tholin, and dotted curve is *k* for $C_2H_6$-$H_2O$ ice tholin.

An alternate approach is to ask what effective optical constants of the haze particles could produce the observed Cthulhu colors. For isotropic single scattering, we would need single scattering albedo, *w*, of 0.30, 0.47, and 0.71 to match I/F at *BLUE*, *RED*, and *NIR* wavelengths respectively. By 1.3 microns, *w* would need to have risen to around 0.96. Assuming a particle size, we can invert the equivalent slab single particle approximation in Hapke's model to say what value of the imaginary part of the refractive index, *k*, would be needed to match *w* at these wavelengths. These values are compared with *k* for various tholins in Fig. 8. The Khare et al. (1984) Titan tholin shows too steep of a gradient through the MVIC wavelength range, as do other Titan tholins (not shown). Tholins produced by irradiation of ices, rather than gases (e.g., Khare et al. 1993, 1994; Materese et al. 2014), show a gentler slope at MVIC wavelengths, more consistent with MVIC colors, but they do not match the steep decline in absorption between MVIC and LEISA wavelengths. Only the Titan tholins drop off quickly enough over that wavelength range. Our assumption of isotropic single scattering may be part of the problem. More plausibly, the single scattering phase function, $P(g)$, should be forward scattering, and also wavelength-dependent. Whether or not the observed spectral behavior can be reconciled with existing tholin optical constants by including such effects is left to future analysis.

## 5.2 Sputnik Planitia

Sputnik Planitia (SP) is a vast, 1,000 by 1,300 km deposit of volatile ice filling what has been interpreted as an ancient impact basin (Moore et al. 2016; McKinnon et al. 2016). From LEISA spectroscopy, the composition of the optically active surface is dominated by $N_2$ ice, with admixtures of CO and $CH_4$ ices (Protopapa et al. 2017; Schmitt et al. 2017). As mapped by White et al. (2017), much of SP's surface features a cellular pattern of irregular polygons a few tens of km across that have been interpreted as the surface expression of convection cells (McKinnon et al. 2016; Trowbridge et al. 2016). Rheological models assuming a pure nitrogen ice composition show that for a plausible internal radiogenic heat flow of 3 mW m$^{-2}$, solid state convection should occur in $N_2$ ice for deposits thicker than ~500 m or so, with convective overturn timescales on the order of $10^5$ to $10^6$ years (McKinnon et al. 2016; Trowbridge et al. 2016). The thickness of the convecting volatile ice deposit is uncertain, but it could be as shallow as ~1 km or as deep as ~10 km, with the depth-to-width aspect ratio of the convection cells being sensitively dependent on boundary conditions and the temperature-dependent viscosity of the ice (McKinnon et al. 2016).

Haze particles deposited on the surface of SP should be transported along with the convective motion until they reach a down-welling zone after some $10^5$ to $10^6$ years. By that time, something like 0.35 to 3.5 mm of material should have accumulated, based on our estimated haze production rate. If the haze particles have appreciable tholin content, as we concluded they must in the previous section, the presence of such a thick coating of them should impart a very strong coloration in MVIC's color images, yet there does not appear to be much of a color gradient from the up-welling centers to the down-welling edges of the cells. A hard color stretch of part of Sputnik is shown in Fig. 9. In some cells, especially toward the east, there is evidence for increasing redness from center to edge, as would be expected for increasing abundance of haze particles accumulating on the older terrain toward the edge of each cell. But in terms of color ratios, the color difference is quite small, as shown in the right panel. Such a subtle color difference could easily be a texture effect, unrelated to haze accumulation. To the north and west of Fig. 9, regions currently undergoing sublimation (Protopapa et al. 2017; Bertrand et al. 2018), some of the upwelling central regions appear darker, though not appreciably redder.

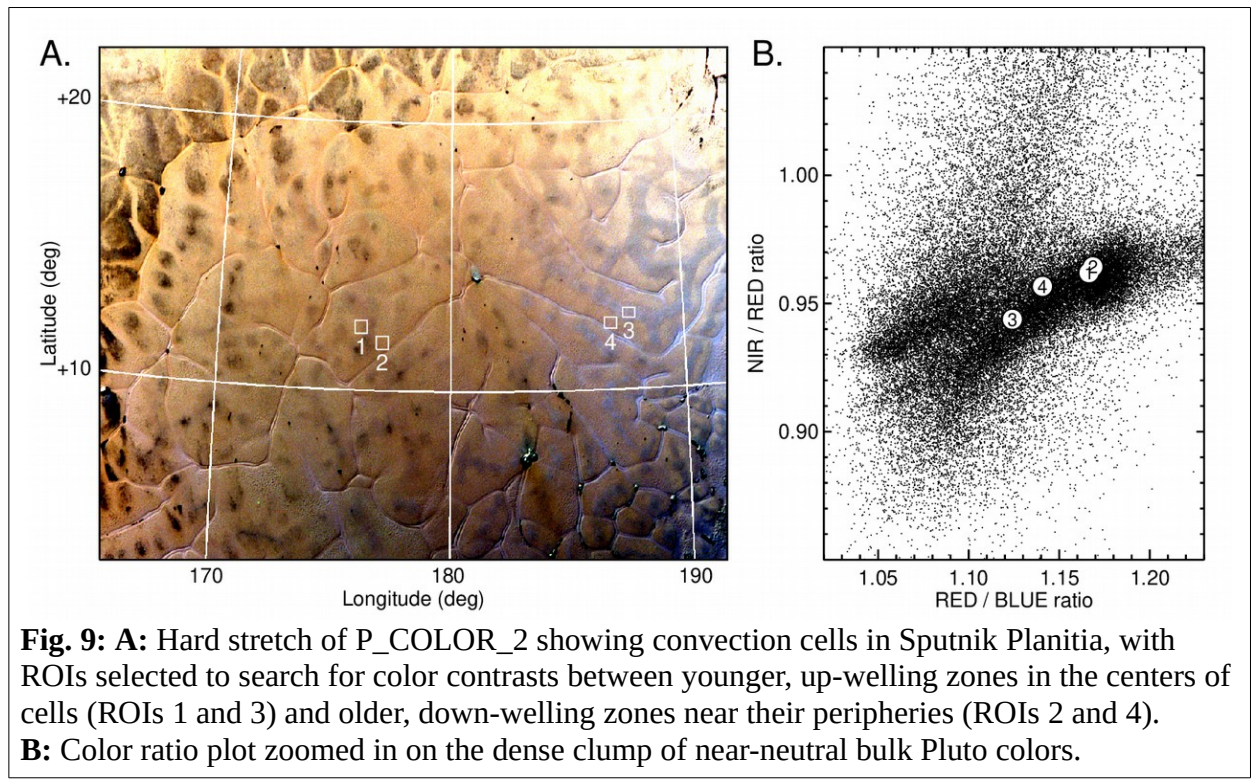

**Fig. 9: A:** Hard stretch of P_COLOR_2 showing convection cells in Sputnik Planitia, with ROIs selected to search for color contrasts between younger, up-welling zones in the centers of cells (ROIs 1 and 3) and older, down-welling zones near their peripheries (ROIs 2 and 4). **B:** Color ratio plot zoomed in on the dense clump of near-neutral bulk Pluto colors.

Several possible mechanisms may explain the absence of stronger color differences between up-welling and down-welling regions in Sputnik Planitia, despite rapid deposition of strongly-pigmented haze particles. Haze particles may be quickly buried by the diurnal (6.4 Earth-day period) or annual (248 Earth-year period) cycles of volatile ice sublimation and condensation. One dimensional global volatile transport models predating the New Horizons flyby had anticipated deposition and sublimation rates as great as a few cm per Earth year (e.g., Spencer et al. 1997; Hansen et al. 2015). Accounting for the actual topography and albedo of Sputnik Planitia, and using a Global Circulation Model (GCM), Forget et al. (2017) and Bertrand et al. (2018) confirm sublimation and deposition rates as high as tens of microns per Pluto day, and locally as high as 5 cm per Pluto year, far in excess of the haze accumulation rate. Additionally, a dark haze particle resting on or near the surface of the volatile $N_2$ ice will absorb more sunlight than adjacent $N_2$ ice does. The resulting additional heating will accelerate $N_2$ sublimation in the immediate vicinity of the haze particle, possibly excavating a cavity into which it could sink (e.g., Kieffer 1990; Grundy & Stansberry 2000; Keiffer et al. 2000; Piqueux et al. 2003; Keifer 2006; Keiffer et al. 2007; Portyankina et al. 2010). If more dark haze particles accumulate in the cavity thus produced, it could grow into a larger pit over time. Complex wind patterns are expected to occur in and around Sputnik due to sublimation and condensation in response to the diurnal insolation cycle and to katabatic downslope flows, according to global circulation models (e.g., Forget et al. 2017). It is not clear whether these winds would be sufficient to prevent haze accumulation in Sputnik, or would concentrate its deposition into certain regions, but depending on the sizes of the haze particles, wind speeds could exceed haze settling speeds and thus transport haze particles over large distances horizontally before they reach the surface (Bertrand & Forget 2017). Winds may also mobilize haze particles after they settle onto the surface (e.g., Telfer et al. 2018), transporting them to nearby pits or fractures, where they would not be so easily seen. Potential dunes mapped by Telfer et al. in northwest Sputnik have high albedos and show $CH_4$ ice absorption so they are clearly not dominated by tholin-rich haze, but even aeolian activity involving $CH_4$ ice particles could be expected to refresh the surface in that region much faster than haze can accumulate.

Wherever haze particles accumulate on SP, though they may be hidden from view through burial or other mechanisms, they will be carried along with the convective flow toward down-welling regions, and ultimately mixed into the bulk ice via the convective churning. If the glacier is 1 km thick and 4 Gyr old, and haze has been accumulating at our estimated rate and mixing into the glacier over that entire time, then about ~1.4% of the glacier's present-day bulk would have its origin in atmospheric haze. If the glacial ice deposit is 10 km thick, reflecting the higher end of estimates (e.g., Trowbridge et al. 2016), then the haze would only have contributed a tenth as much, fractionally. Depending on the tholin content of the haze, even such a small admixture should be enough to alter the coloration of the bulk ice, since tholin is so strongly pigmented and $N_2$ ice is highly transparent. Even a fraction of a percent of haze should make exposed slab ice look dark rather than bright. Indeed, this may account for the dark floors of sublimation pits in southern Sputnik Planitia (e.g., White et al. 2017) as well as the darker ice along the northwest margin of Sputnik where climate models suggest sublimation has recently been most active (Bertrand et al. 2018). Whether such quantities of heavier hydrocarbons and nitriles would significantly alter the phase behavior or rheology of $N_2$:CO:$CH_4$ ice is unclear, but laboratory studies are clearly called for. Though their contribution may be volumetrically small, relatively non-volatile particles could act as pinning points and thus influence the rheologically important grain size distribution.

## 5.3 Lowell Regio

Pluto has a variety of regions that have been identified as being especially rich in $CH_4$ ice. These are highlighted in maps of the equivalent width of absorption in the MVIC *CH4* filter covering 860-910 nm (Grundy et al. 2016; Earle et al. 2018) as well as in LEISA maps (Protopapa et al. 2017; Schmitt et al. 2017). $CH_4$-rich regions include Lowell Regio (the north polar zone), the bladed terrain of Tartarus Dorsa, various bright mid-latitude crater rims and scarps, and the ridge crest of Enrique Montes in eastern Cthulhu.

This section will focus on Lowell Regio. Parts of this high-latitude region show geomorphological evidence for kilometer thick mantling of underlying topography (Howard et al. 2017). Such thick ice deposits would require many Pluto seasons to accumulate and so could perhaps be related to the ~3 Myr cycle of mega-seasons (Dobrovolskis et al. 1997; Earle & Binzel 2015; Bertrand et al. 2018), or maybe much longer, as suggested by the existence of a few small impact craters. Bertrand et al. (manuscript in preparation) estimate km thick $CH_4$-dominated polar deposits would require time scales of order 100 Myr to form. A deposit that took 1 Myr to form could be expected to incorporate 3.5 mm of haze, assuming steady state haze production and deposition. For a 1 km thick deposit, that would be 3.5 parts per million of its bulk volume, which sounds negligible, but if it arrives in the form of 1 μm haze particles, they would add up to ~3.5 million particles per cubic centimeter of deposited ice. They could alter the bulk ice's rheological behavior by pinning grain boundaries. Or, some of the haze constituents could dissolve into the ice (solubilities of the relevant hydrocarbons and nitriles in ices of either $CH_4$ or $N_2$ are unknown). If the thick ice deposits accumulate over 100 Myr, the fraction of haze incorporated would be 100 × greater, again assuming a constant production rate.

Another possibility is that the ice accumulates relatively quickly, as an initially $N_2$-dominated winter seasonal ice, but the more volatile $N_2$ is then distilled out via sublimation loss the following summer, leaving a thin residue of the less volatile $CH_4$ ice, along with whatever payload of haze particles had been deposited along with the condensing volatile ices (possibly modified by partial dissolution into the $N_2$ ice prior to its sublimation). Over many seasonal deposition and sublimation cycles, a thick, but finely-layered, $CH_4$-dominated deposit that includes abundant haze particles could accumulate. After the volatile $N_2$ is distilled off during summers, the remaining $CH_4$ might get warm enough for sintering and grain growth. Due to escape of radiogenic heat from Pluto's interior, this material would also get warmer as it is buried more deeply, and the combination of increasing temperature and overburden pressure could lead to metamorphism and/or partial melting.

It may be possible to place a weak constraint on the age or deposition rate of the polar $CH_4$ deposits from the fact that their albedos are generally high, with *RED* filter I/F values in the 0.6 to 0.7 range in the P_COLOR_2 observation. Such high albedos are consistent with deposition and subsequent erosion too fast for strong radiolytic darkening by cosmic rays. That constraint would be much tighter if there were regular episodes of atmospheric collapse, allowing Ly α and perhaps solar wind to reach the surface, but for purposes of this paper, we assume that does not happen and cosmic rays are the primary drivers of radiolysis at the surface. Deposition as slow as ~5 mm in 100 Myr would enable appreciable cosmic ray processing during deposition, from the estimates of Strazzulla et al. (1984), so Pluto's Lowell Regio deposits must have formed much faster than that to avoid radiolytic darkening. But such slow deposition is also inconsistent with the much more rapid rate of haze accumulation, so little is gained from excluding low deposition rates on the basis of cosmic ray processing.

Another mechanism that may help to limit the coloration from haze settling on Lowell Regio is the enhanced absorption of sunlight by dark haze particles leading to local sublimation that causes them to subside below the optically active surface and thus disappear from view (Kieffer 1990; Grundy & Stansberry 2000; Keiffer et al. 2000; Piqueux et al. 2003; Keifer 2006; Keiffer et al. 2007; Portyankina et al. 2010). This mechanism would operate much more slowly in Lowell Regio's less volatile $CH_4$ ice than it would in the $N_2$-dominated ice of Sputnik Planitia, but it could still be important.

A distinctive feature of the MVIC *BLUE-RED-NIR* images of Lowell Regio is its golden hue, especially in the uplands. On a color ratio plot (Olkin et al. 2017), this "gold" color follows a different mixing line from the one between bulk Pluto colors and Cthulhu (see Fig. 10).

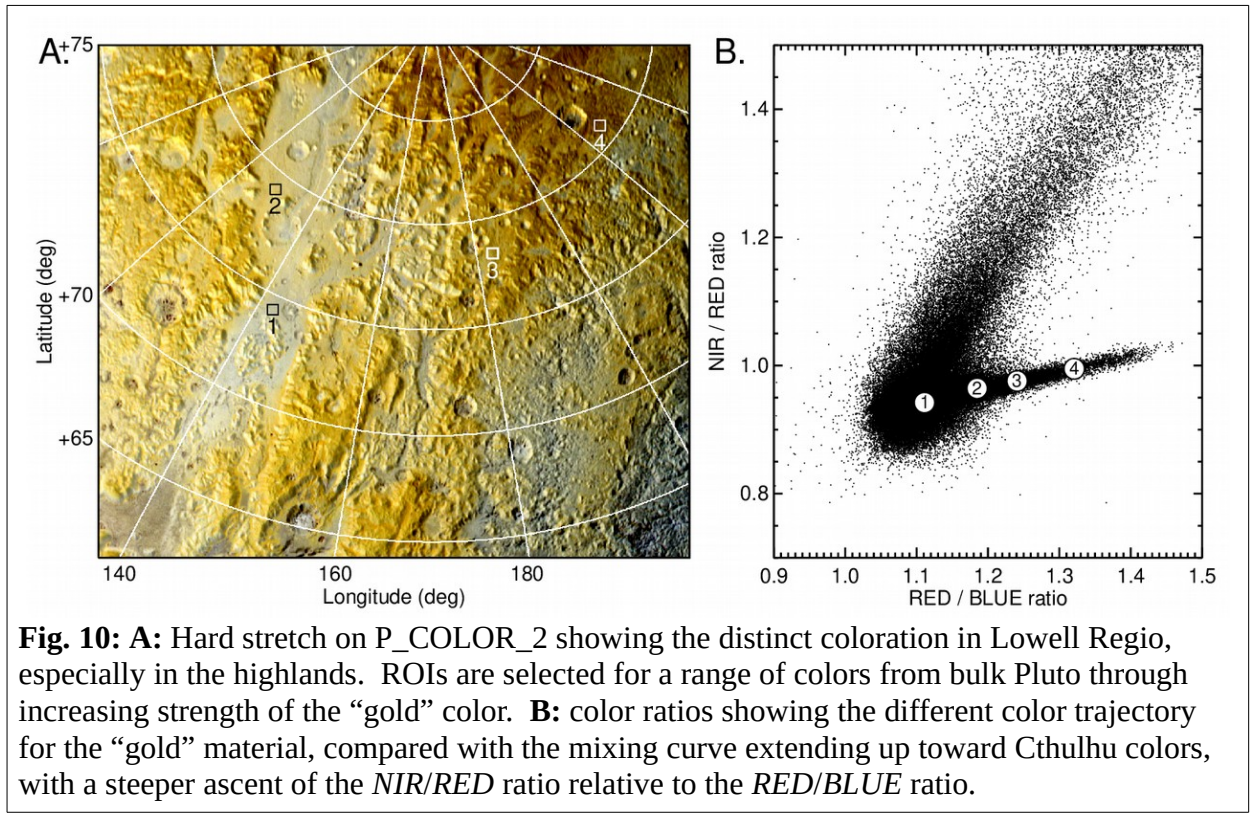

**Fig. 10: A:** Hard stretch on P_COLOR_2 showing the distinct coloration in Lowell Regio, especially in the highlands. ROIs are selected for a range of colors from bulk Pluto through increasing strength of the "gold" color. **B:** color ratios showing the different color trajectory for the "gold" material, compared with the mixing curve extending up toward Cthulhu colors, with a steeper ascent of the *NIR*/*RED* ratio relative to the *RED*/*BLUE* ratio.

A key question is whether it is possible to reconcile this very different trajectory in color space with that of Cthulhu, using the same model haze particles as colorants in both regions. We tried a variety of models using Titan tholin optical constants as well as the synthetic Cthulhu tholin optical constants from Fig. 8, but in all cases, we were unable to match Lowell Regio's steep spectral slope between MVIC's *BLUE* and *RED* filters without exceeding the much shallower spectral slope between the *RED* and *NIR* filters. Noting that $CH_4$ ice is abundant in this region, and that $CH_4$ ice has appreciable absorption in MVIC's *NIR* filter, we also tried models where we used $CH_4$ absorption to suppress the spectral slope between *RED* and *NIR* channels. It was possible to match the "gold" color using such a configuration, but not the high albedo. A comparison of the spatial distribution of the strong $CH_4$ absorption and the "gold" color (Fig. 11) shows that the two are not perfectly correlated, as would be expected if it was the combination of $CH_4$ plus haze particles that produced Lowell Regio's distinctive coloration. In particular, the "gold" color can be seen at the base of some steep slopes, in addition to the uplands, as if it erodes out of the uplands and moves downslope, whereas the strong $CH_4$ absorption appears to be confined to the uplands.

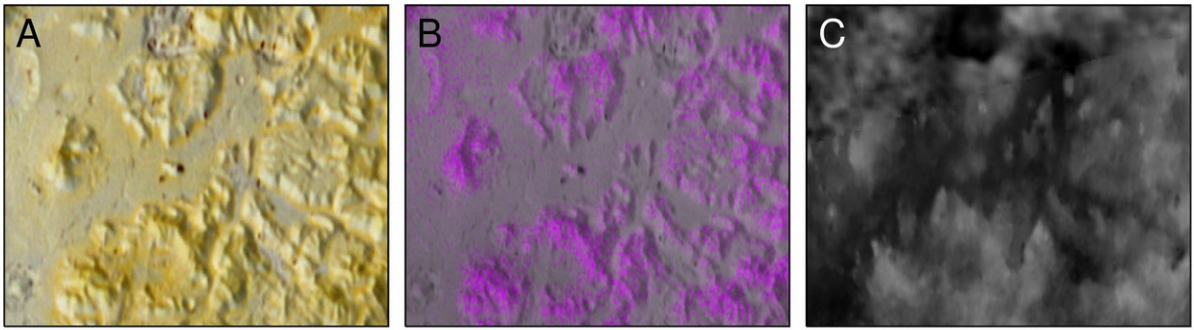

**Fig. 11:** Zoom on a 150 km wide region near 154° E, 78° N, showing the spatial distribution of the "gold" material. The $CH_4$ absorption tends to be associated with uplands, while the distribution of the "gold" material is more complicated, being found both in uplands and near the bases of steep slopes. **A:** MVIC *BLUE-RED-NIR* color. **B:** MVIC $CH_4$ equivalent width map painted in magenta on top of the panchromatic base map. **C:** Elevation from Schenk et al. (2018), stretched from −2 to +4 km altitude relative to the 1,188.3 km mean radius (Nimmo et al. 2017).

## 6. Discussion

Our inability to match Cthulhu and Lowell Regio colors with models that incorporate a common model haze particle presents a conundrum. Haze forms in Pluto's atmosphere and settles out globally and at a relatively rapid pace, so why is it not coloring the surface more uniformly? Here we suggest several possible explanations, but we are not able to determine which, if any, of these is correct.

One idea is that the "gold" color of Lowell Regio is due not to haze particles, but to *in situ* radiolysis or photolysis of local ices. The main problem with this idea is that it is difficult to understand how a deposit old enough to accumulate a substantial radiation dose could avoid being completely buried by new haze deposition, considering the screening effect of Pluto's atmosphere. And local geological activity (e.g., mass wasting) could easily overwhelm both haze deposition and radiolysis by the energetic radiation that does make it through Pluto's atmosphere. A possible solution could be that the radiation coincides with periods of atmospheric collapse that permit more radiation to reach the surface at the same time as they interrupt atmospheric haze production. If the atmosphere seasonally collapses, and it is deposited mostly on the winter pole, then the UV that reaches the surface there could be processing a different mix of ices (mostly $N_2$ with smaller amounts of $CH_4$ and CO) compared to the mix of nitriles and hydrocarbons from prior haze fallout that is being irradiated on Cthulhu. Whether radiolysis of these different chemical feedstocks can account for the distinct observed colors calls for further laboratory work, though it seems plausible. A key difference is that in an $N_2$-dominated ice, the combination of $CH_4$ radiolysis products into larger hydrocarbon molecules requires their diffusion through the $N_2$ ice lattice. The net result may be smaller radiolytic product molecules, perhaps with less loss of H, and/or greater incorporation of N (e.g., Wu et al. 2012; Materese et al. 2014). Different radiation sources might also be relevant. The winter pole will tend to receive less solar wind, and the UV it receives is limited to the solar Ly α resonantly scattered by the interplanetary medium and emission from nearby stars. Equatorial regions like Cthulhu are directly irradiated by the broader spectrum of solar UV. But rapid radiolysis or photolysis of surface ices requires a seasonal atmospheric collapse to allow much more radiation to reach Pluto's surface than arrives through highly penetrating cosmic rays. Long term GCM models argue that this does not occur (e.g., Bertrand & Forget 2016; Forget et al. 2017; Bertrand et al. 2018). Sporadic astrophysical events offer alternate ways of increasing the dose of energetic radiation at Pluto's surface, but the radiolytic products from such an event would soon be buried under subsequent haze fallout.

Another possibility involves seasonally differing haze compositions, resulting from seasonally changing atmospheric composition or structure. Different haze could get incorporated into different regions, if it preferentially accumulates wherever seasonal condensation/deposition of volatiles is occurring at the time that the haze is produced (e.g., Bertrand & Forget 2017). Evidence from stellar occultations suggests that Pluto's haze opacity varies on relatively short timescales, perhaps being influenced by the solar cycle (e.g., E.F. Young et al. 2018). Ly α only varies by a factor of ~2 over the solar cycle, so a strong variability in haze opacity correlated to the solar cycle would suggest a key role in haze production for the more energetic UV photons that vary more strongly with the solar cycle. However, temporal variability of the haze opacity could also result from seasonal evolution of Pluto's atmospheric density and circulation, without requiring changes in its production rate (e.g., Bertrand & Forget 2017). Spatially variable haze production is possible, too. Over the poles, direct solar Ly α is incident from a higher annual average elevation angle and thus penetrates deeper into the atmosphere on average than it does over the equator, driving photolysis at a lower average altitude. At the time of the flyby, lower altitudes featured more $N_2$ relative to $CH_4$ and fewer charged particles.

Another possible way that different sorts of haze particles could be delivered to different regions is through the effect of thermal processing on descent. The extent of the lower atmosphere processing could vary seasonally and as a function of latitude. Haze particles enlarged by a payload of lighter hydrocarbons (perhaps favored over the un-illuminated winter pole) might settle out comparatively quickly, while particles that are warmed by sunlight over the summer pole might accrete less hydrocarbons, settle more slowly, and be more dispersed by winds as they settle. This scenario may be testable by studying the lowest altitude haze band where the influence of seasonal near-surface temperature variations could be greatest.

Even if the haze arriving at Pluto's surface is regionally and temporally uniform, a distribution of haze particle characteristics such as size or tholin content could result in distinct appearances in different environments. For instance, in Cthulhu, the entire distribution of haze particles could be expected to remain visible, unless one type was more readily mobilized by aeolian transport than another. This behavior contrasts with Lowell Regio, where the most strongly-absorbing subset of haze particles could preferentially sink below view into the volatile $CH_4$ ice, leaving the subset of particles that absorb the least sunlight as the dominant colorant at the surface.

Finally, it is possible that the "gold" material seen at Lowell Regio is what Pluto's haze looks like after having been buried in $CH_4$ ice. Seeing such exhumed material at the surface today would require substantial sublimation loss of polar ice to expose it again after having been buried, but perhaps that is consistent with the northern pole having been the most recent one where summer solstice coincided with perihelion (around ~1 Myr ago, Earle & Binzel 2015). Little is known about how Pluto's haze materials would behave under relevant temperature and pressure conditions, but the mobility of molecules in weakly-bound, low temperature solids can be significant, and below the surface, warmer temperatures and higher pressures would facilitate metamorphosis. Much more laboratory work is needed to explore how potential haze materials might behave under such conditions. An extension of this possibility is that haze particles could undergo more substantial chemical evolution after their arrival at Pluto's surface, particularly if they arrive there with residual dangling bonds or immobilized radicals from their formation. With lower surface temperatures, chemical evolution can be expected to progress much more slowly on the surface than it does in the upper atmosphere, but this obstacle may be overcome by the fact that residence times on the surface are substantially longer. Potential types of reactions that might occur include incorporation of locally condensed volatile ices such as $CH_4$ or CO into the macromolecular structure of the tholins. This sort of thing might occur in Lowell Regio or Sputnik Planitia, where haze particles are dispersed in abundant volatile ices. Reactions involving these ices might integrate more hydrogen or oxygen into the tholin, possibly modifying its aromatic/aliphatic ratio and its color. Even in places like Cthulhu where other materials may be less available, it is possible to envision further reactions progressing within the tholin itself. Such reactions could alter its structure over time, perhaps reducing the hydrogen abundance and increasing the prevalence of aromatic rings.

Bladed terrain (e.g., Moore et al. 2016, 2018; Moores et al. 2017) presents a related challenge, and perhaps also a clue. It and Lowell Regio are both $CH_4$-dominated regions (at least for their uppermost surfaces sampled by reflectance spectroscopy), but bladed terrain does not show the "gold" color of Lowell Regio in the *BLUE-RED-NIR* images from MVIC. The color of bladed terrain is more of a greenish hue in those images. Adding additional $CH_4$ absorption to the gold material may be a way to produce colors like that, since it adds additional absorption in the *NIR* channel. The most obvious environmental differences between bladed terrain and Lowell Regio are related to latitude, altitude, and thermal history, suggesting a link to the distinct seasonal thermal histories experienced by different latitudes on Pluto (Binzel et al. 2017): Lowell Regio's "gold" material is confined to high latitudes (e.g., Olkin et al. 2017), while bladed terrain is only seen at low latitudes (e.g., Olkin et al 2017; Moore et al. 2018). Volatile ices are sublimating from Lowell Regio during the present season, whereas they are likely to be condensing on the Bladed terrain. Perhaps the "gold" color is simply a consequence of distinct textures associated with sublimation, while the greener shade of bladed terrain is what that same combination of materials looks during times when condensation prevails.

Ultimately, this analysis appears to be indicating that the uniformitarian assumption of steady state haze production of a material that is essentially inert on Pluto's surface is inconsistent with the observed diversity of Pluto's surface environments. Either the haze production and delivery are not spatially or temporally uniform, or its evolution after arrival at the surface follows divergent trajectories in different environments, or some combination of these things must be true.

## Acknowledgments

This work was supported in part by NASA's New Horizons project. We gratefully thank NASA and the entire New Horizons team for their hard work leading to a spectacularly successful Pluto system encounter. B. Schmitt and E. Quirico acknowledge France's Centre National d'Etudes Spatiales (CNES) for its financial support through its "Système Solaire" program. Finally, we thank the free and open source software communities for empowering us with key tools used to complete this project, notably Linux, the GNU tools, LibreOffice, MariaDB, ISIS3, QGIS, Evolution, Python, the Astronomy User's Library, and FVWM.

## References

Bagenal, F., and 25 co-authors 2016. Pluto's interaction with its space environment: Solar wind, energetic particles, and dust. *Science* **351,** 1282.

Bertrand, T., and F. Forget 2016. Observed glacier and volatile distribution on Pluto from atmosphere-topography processes. *Nature* **540,** 86-89.

Bertrand, T., and F. Forget 2017. 3D modeling of organic haze in Pluto's atmosphere. *Icarus* **287,** 72-86.

Bertrand, T., and 13 co-authors 2018. The nitrogen cycles on Pluto over astronomical timescales. *Icarus* **309,** 277-296.

Binzel, R.P., and 12 co-authors 2017. Climate zones on Pluto and Charon. *Icarus* **287,** 30-36.

Buratti, B.J. 1985. Application of a radiative transfer model to bright icy satellites. *Icarus* **61,** 208-217.

Buratti, B.J., and J.A. Mosher 1995. The dark side of Iapetus: New evidence for an exogenous origin. *Icarus* **115,** 219-227.

Buratti, B.J., and J. Veverka 1985.  Photometry of rough planetary surfaces: the role of multiple scattering.  *Icarus* **64,** 320-328.

Buratti, B.J. et al. 2006.  Titan: preliminary results on surface properties and photometry from VIMS observations of the early flybys.  *Planet. & Space Sci.* **54,** 1498-1509.

Buratti, B.J., and 16 co-authors 2017.  Global albedos of Pluto and Charon from LORRI New Horizons observations.  *Icarus* **287,** 207-217.

Chandrasekhar, S. 1960.  *Radiative Transfer*.  Dover Press, New York.

Cheng, A.F., and 15 co-authors 2008.  Long-Range Reconnaissance Imager on New Horizons.  *Space Sci. Rev.* **140,** 189-215.

Cheng, A.F., and 13 co-authors 2017.  Haze in Pluto's atmosphere.  *Icarus* **290,** 112-133.

Cook, J.C., and 27 co-authors 2018.  The distribution of $H_2O$, $CH_3OH$, and hydrocarbon-ices on Pluto: Analysis of New Horizons spectral images.  *Icarus* (submitted).

Cooper, J.F., E.R. Christian, J.D. Richardson, and C. Wang 2003.  Proton irradiation of Centaur, Kuiper belt, and Öort cloud objects at plasma to cosmic ray energy.  *Earth, Moon, & Planets* **92,** 261-277.

Cravens, T.E., and D.F. Strobel 2015.  Pluto's solar wind interaction: Collisional effects.  *Icarus* **246,** 303-309.

Denk, T., and 10 co-authors 2009.  Iapetus: Unique surface properties and a global color dichotomy from Cassini imaging.  *Science* **327,** 435-439.

Dobrovolskis, A.R., S.J. Peale, and A.W. Harris 1997.  Dynamics of the Pluto-Charon binary.  In: S.A. Stern, D.J. Tholen (Eds.), *Pluto and Charon*, University of Arizona Press, Tucson, 159-190.

Domingue, D.L., G.W. Lockwood, and D.T. Thompson 1995.  Surface textural properties of icy satellites: A comparison between Europa and Rhea.  *Icarus* **115,** 228-249.

Earle, A.M., and R.P. Binzel 2015.  Pluto's insolation history: Latitudinal variations and effects on atmospheric pressure.  *Icarus* **250,** 405-412.

Earle, A.M., R.P. Binzel, L.A. Young, S.A. Stern, K. Ennico, W. Grundy, C.B. Olkin, H.A. Weaver, and the New Horizons Geology and Geophysics Imaging Team 2017.  Long-term surface temperature modeling of Pluto.  *Icarus* **287,** 37-46.

Earle, A.M., and 18 co-authors 2018.  Methane distribution on Pluto as mapped by the New Horizons Ralph/MVIC instrument.  *Icarus* (in press).

Eluszkiewicz, J., J. Leliwa-Kopystyński, and K.J. Kossacki 1998.  Metamorphism of solar system ices.  In: B. Schmitt, C. de Bergh, M. Festou (Eds.), *Solar System Ices,* Kluwer Academic Publishers, Boston, 119-138.

Forget, F., F. Bertrand, M. Vangvichith, J. Leconte, E. Millour, and E. Lellouch 2017.  A post-New Horizons global climate model of Pluto including the $N_2$, $CH_4$, and CO cycles.  *Icarus* **287,** 54-71.

Fray, N., and B. Schmitt 2009.  Sublimation of ices of astrophysical interest: A bibliographic review.  *Planet. Space Sci.* **57,** 2053-2080.

Gao, P., and 13 co-authors 2017.  Constraints on the microphysics of Pluto's photochemical haze from New Horizons observations.  *Icarus* **287,** 116-123.

Garnett, J.C.M. 1904.  Colours in metal glasses and in metallic films.  *Phil. Trans. Roy. Soc.* **203,** 385-420.

Gladstone, G.R., W.R. Pryor, and S.A. Stern 2015.  Lyα@Pluto.  *Icarus* **246,** 279-284.

Gladstone, G.R., and 33 co-authors 2016.  The atmosphere of Pluto as observed by New Horizons.  *Science* **351,** 1280.

Grundy, W.M., and J.A. Stansberry 2000.  Solar gardening and the seasonal evolution of nitrogen ice on Triton and Pluto.  *Icarus* **148,** 340-346.

Grundy, W.M., and 33 co-authors 2016a.  Surface compositions across Pluto and Charon.  *Science* **351,** 1283.

Grundy, W.M., and 39 co-authors 2016b.  Formation of Charon's red poles from seasonally cold-trapped volatiles.  *Nature* **539,** 65-68.

Hamilton, D.P., S.A. Stern, J.M. Moore, and L.A. Young 2016.  The rapid formation of Sputnik Planitia early in Pluto's history.  *Nature* **540,** 97-99.

Hanley, J., L. Pearce, G. Thompson, W. Grundy, H. Roe, G. Lindberg, S. Dustrud, D. Trilling, and S. Tegler 2017.  Methane, ethane, and nitrogen liquid stability on Titan.  *Lunar & Planetary Science Conference* abstract 1686.

Hansen, C.J., D.A. Paige, and L.A. Young 2015.  Pluto's climate modeled with new observational constraints.  *Icarus* **246,** 183-191.

Hapke, B. 1984.  Bidirectional reflectance spectroscopy. 3. Correction for macroscopic roughness.  *Icarus* **59,** 41-59.

Hapke, B. 1993.  *Theory of reflectance and emittance spectroscopy*.  Cambridge University Press, New York.

Helfenstein, P., and J. Veverka 1987. Photometric properties of lunar terrains derived from Hapke's equation. *Icarus* **72,** 342-357.

Hinson, D.P., and 15 co-authors 2017. Radio occultation measurements of Pluto's neutral atmosphere with New Horizons. *Icarus* **290,** 96-111.

Holler, B.J., L.A. Young, W.M. Grundy, C.B. Olkin, and J.C. Cook 2014. Evidence for longitudinal variability of ethane ice on the surface of Pluto. *Icarus* **243,** 104-110.

Howett, C.J.A., and 37 co-authors 2017. Inflight radiometric calibration of New Horizons' Multispectral Visible Imaging Camera (MVIC). *Icarus* **287,** 140-151.

Imanaka, H., B.N. Khare, J.E. Elsila, E.L.O. Bakes, C.P. McKay, D.P. Cruikshank, S. Sugita, T. Matsui, and R.N. Zare 2004. Laboratory experiments of Titan tholin formed in cold plasma at various pressures: Implications for nitrogen-containing polycyclic aromatic compounds in Titan haze. *Icarus* **168,** 344-366.

Johnson, R.E. 1990. *Energetic charged-particle interactions with atmospheres and surfaces*. Springer-Verlag, New York.

Kieffer, H.H. 1990. $H_2O$ grain size and the amount of dust in Mars' residual north polar cap. *J. Geophys. Res.* **95,** 1481-1493.

Kieffer, H.H., T.N. Titus, K.F. Mullins, and P.R. Christensen 2000. Mars south polar spring and summer behavior observed by TES: Seasonal cap evolution controlled by frost grain size. *J. Geophys. Res.* **105,** 9653-9699.

Kieffer, H.H., P.R. Christensen, and T.N. Titus 2006. $CO_2$ jets formed by sublimation beneath translucent slab ice in Mars' seasonal south polar ice cap. *Nature* **442,** 793-796.

Kieffer, H.H. 2007. Cold jets in the martian polar caps. *J. Geophys. Res.* **112,** E08005.1-15.

Kim, Y.S., and R.I. Kaiser 2012. Electron irradiation of Kuiper belt surface ices: Ternary $N_2$-$CH_4$-CO mixtures as a case study. *Astrophys. J.* **758,** 37.1-6.

Khare, B.N., C. Sagan, E.T. Arakawa, F. Suits, T.A. Callcott, and M.W. Williams 1984. Optical constants of organic tholins produced in a simulated Titanian atmosphere: From soft X-ray to microwave frequencies. *Icarus* **60,** 127-137.

Khare, B.N., W.R. Thompson, L. Cheng, C. Chyba, C. Sagan, E.T. Arakawa, C. Meisse, and P.S. Tuminello 1993. Production and optical constants of ice tholin from charged particle irradiation of (1:6) $C_2H_6$/$H_2O$ at 77 K. *Icarus* **103,** 290-300.

Khare, B.N., C. Sagan, M. Heinrich, W.R. Thompson, E.T. Arakawa, P.S. Tuminello, and M. Clark 1994. Optical constants of Triton tholin: Preliminary results. *Bull. Am. Astron. Soc.* **26,** 1176-1177 (abstract).

Lavvas, P., and 11 co-authors 2013. Aerosol growth in Titan's ionosphere. *Publ. Nat. Acad. Sci.* **110,** 2729-2734.

Lee, J.S., B.J. Buratti, M. Hicks, and J.A. Mosher 2010. The roughness of the dark side of Iapetus from the 2004 to 2005 flyby. *Icarus* **206,** 623-630.

Lisse, C.M., and 18 co-authors 2017. The puzzling detection of x-rays from Pluto by Chandra. *Icarus* **287,** 103-109.

Luspay-Kuti, A., K. Mandt, K.L. Jessup, J.A. Kammer, V. Hue, M. Hamel, and R. Filwett 2017. Photochemistry on Pluto – I. Hydrocarbons and aerosols. *Mon. Not. R. Astron. Soc.* **472,** 104-117.

Materese, C.K., D.P. Cruikshank, S.A. Sandford, H. Imanaka, M. Nuevo, and D.W. White 2014. Ice chemistry on outer solar system bodies: Carboxylic acids, nitriles, and urea detected in refractory residues produced from the UV photolysis of $N_2$:$CH_4$:CO-containing ices. *Astrophys. J.* **788,** 111.1-11.

McKinnon, W.B., and 14 co-authors 2016. Convection in a volatile nitrogen-ice-rich layer drives Pluto's geological vigour. *Nature* **534,** 82-85.

Mewaldt, R.A., C.M.S. Cohen, G.M. Mason, D.K. Haggerty, and M.I. Desai 2007. Long-term fluences of solar energetic particles from H to Fe. *Space Sci. Rev.* **130,** 323-328.

Moore, J.M., and 40 co-authors 2016. The geology of Pluto and Charon through the eyes of New Horizons. *Science* **351,** 1284-1293.

Moore, J.M., and 15 co-authors 2018. Bladed terrain on Pluto: Possible origins and evolution. *Icarus* **300,** 129-144.

Moores, J.E., C.L. Smith, A.D. Toigo, and S.D. Guzewich 2017. Penitentes as the origin of the bladed terrain of Tartarus Dorsa on Pluto. *Nature* **541,** 188-190.

Nimmo, F., and 16 co-authors 2017. Mean radius and shape of Pluto and Charon from New Horizons images. *Icarus* **287,** 12-29.

Olkin, C.B., and 24 co-authors 2017. The color of Pluto from New Horizons. *Astron. J.* **154,** 258.1-13.

Padovani, M., D. Galli, and A.E. Glassgold 2009. Cosmic-ray ionization of molecular clouds. *Astron. & Astrophys.* **501,** 619-631.

Peterson, J.G., E. Birath, B. Carcich, and A. Harch 2013. Closing the uplink/downlink loop on the New Horizons mission to Pluto. *Proceedings of Aerospace Conf., 2013 IEEE,* 290.1-8.

Piqueux, S., S. Byrne, and M.I. Richardson 2003.  Sublimation of Mars's southern seasonal CO$_2$ ice cap and the formation of spiders.  *J. Geophys. Res.* **108,** E85084.1-9.

Poppe, A.R. 2015.  Interplanetary dust influx to the Pluto-Charon system.  *Icarus* **246,** 352-359.

Porter, S.B., and W.M. Grundy 2015.  Ejecta transfer in the Pluto system.  *Icarus* **246,** 360-368.

Portyankina, G., W.J. Markiewicz, N. Thomas, C.J. Hansen, and M. Milazzo 2010.  HiRISE observations of gas sublimation-driven activity in Mars' southern polar regions: III. Models of processes involving translucent ice.  *Icarus* **205,** 311-320.

Protopapa, S., and 22 co-authors 2017.  Pluto's global surface composition through pixel-by-pixel Hapke modeling of New Horizons Ralph/LEISA data.  *Icarus* **287,** 218-228.

Quirico, E., and 14 co-authors 2008.  New experimental constraints on the composition and structure of tholins.  *Icarus* **198,** 218-231.

Ramirez, S.I., P. Coll, A. da Silva, R. Navarro-González, J. Lafait, and F. Raulin 2002. Complex refractive index of Titan's aerosol analogues in the 200-900 nm domain.  *Icarus* **156,** 515-529.

Rao, C.N.R. 1975.  *Ultra-Violet and Visible Spectroscopy: Chemical Applications*. Butterworths Publishers Inc., Boston.

Reuter, D.C., and 19 co-authors 2008.  Ralph: A visible/infrared imager for the New Horizons Pluto/Kuiper belt mission.  *Space Sci. Rev.* **140,** 129-154.

Robbins, S.J., and 28 co-authors 2017.  Craters of the Pluto-Charon system.  *Icarus* **287,** 187-206.

Schenk, P.M., and 16 co-authors 2018.  Global- and regional-scale cartography and topography of Pluto from New Horizons.  *Icarus* (submitted).

Schmitt, B., and 27 co-authors 2017.  Physical state and distribution of materials at the surface of Pluto from New Horizons LEISA imaging spectrometer.  *Icarus* **287,** 229-260.

Sciamma-O'Brien, E., P.R. Dahoo, E. Hadamcik, N. Carrasco, E. Quirico, C. Szopa, and G. Cernogora 2012.  Optical constants from 370 nm to 900 nm of Titan tholins produced in a low pressure RF plasma discharge.  *Icarus* **218,** 356-363.

Sekine, Y., H. Genda, S. Kamata, and T. Funatsu 2017.  The Charon-forming giant impact as a source of Pluto's dark equatorial regions.  *Nature Astron.* **1,** 0031.1-6.

Spencer, J.R., J.A. Stansberry, L.M. Trafton, E.F. Young, R.P. Binzel, and S.K. Croft 1997. Volatile transport, seasonal cycles, and atmospheric dynamics on Pluto. In: S.A. Stern, D.J. Tholen (Eds.), *Pluto and Charon*, University of Arizona Press, Tucson, 435-473.

Stern, S.A., and 150 co-authors 2015. The Pluto system: Initial results from its exploration by New Horizons. *Science* **350,** 292.

Stern, S.A., R.P. Binzel, A.M. Earle, K.N. Singer, L.A. Young, H.A. Weaver, C.B. Olkin, K. Ennico, J.M. Moore, W.B. McKinnon, J.R. Spencer, and the New Horizons Geology and Geophysics and Atmospheres teams 2017. Past epochs of significantly higher pressure atmospheres on Pluto. *Icarus* **287,** 47-53.

Strazzulla, G., L. Calcagno, and G. Foti 1984. Build up of carbonaceous material by fast protons on Pluto and Triton. *Astron. & Astrophys.* **140,** 441-444.

Strobel, D.F., and X. Zhu 2017. Comparative planetary nitrogen atmospheres: Density and thermal structures of Pluto and Triton. *Icarus* **291,** 55-64.

Tamayo, D., J.A. Burns, D.P. Hamilton, and M.M. Hedman 2011. Finding the trigger to Iapetus' odd global albedo pattern: Dynamics of dust from Saturn's irregular satellites. *Icarus* **215,** 260-278.

Telfer, M.W., and 18 co-authors 2018. Dunes on Pluto. *Science* **360,** 992-997.

Thompson, W.R., B.G.J.P.T. Murray, B.N. Khare, and C. Sagan 1987. Coloration and darkening of methane clathrate and other ices by charged particle irradiation: Applications to the outer solar system. *J. Geophys. Res.* **92,** 14933-14947.

Trowbridge, A.J., H.J. Melosh, J.K. Steckloff, and A.M. Freed 2016. Vigorous convection as the explanation for Pluto's polygonal terrain. *Nature* **534,** 79-81.

Verbiscer, A.J., M.F. Skrutskie, and D.P. Hamilton 2009. Saturn's largest ring. *Nature* **461,** 1098-1100.

Verbiscer, A.J., and J. Veverka 1992. Mimas: Photometric roughness and albedo map. *Icarus* **99,** 63-69.

Vuitton, V., P. Lavvas, R.V. Yelle, M. Galand, A. Wellbrock, G.R. Lewis, A.J. Coates, and J.E. Wahlund 2009a. Negative ion chemistry in Titan's upper atmosphere. *Planet. & Space Sci.* **57,** 1558-1571.

Vuitton, V., B.N. Tran, P.D. Persans, and J.P. Ferris 2009b. Determination of the complex refractive indices of Titan haze analogs using photothermal deflection spectroscopy. *Icarus* **203,** 663-671.

White, O.L., and 23 co-authors 2017. Geological mapping of Sputnik Planitia on Pluto. *Icarus* **287,** 261-286.

Wong, M.L., and 13 co-authors 2017. The photochemistry of Pluto's atmosphere as illuminated by New Horizons. *Icarus* **287,** 110-115.

Wu, Y.J., C.Y.R. Wu, S.L. Chou, M.Y. Lin, H.C. Lu, J.I. Lo, and B.M. Cheng 2012. Spectra and photolysis of pure nitrogen and methane dispersed in solid nitrogen with vacuum-ultraviolet light. *Astrophys. J.* **746,** 175.1-11.

Young, E.F., and 11 co-authors 2018. Pluto's evolving haze opacity from 2002-2015: Correlation to solar activity. *Icarus* (submitted).

Young, L.A., and 25 co-authors 2018. Structure and composition of Pluto's atmosphere from the New Horizons solar ultraviolet occultation. *Icarus* **300,** 174-199.

Zhang, X., D.F. Strobel, and H. Imanaka 2017. Haze heats Pluto's atmosphere yet explains its cold temperature. *Nature* **551,** 352-355.